\documentclass[10pt,journal,comsoc]{IEEEtran}
\usepackage{url}
\usepackage{caption}
\usepackage[noadjust]{cite}
\usepackage{graphicx}
\usepackage{multirow}
\usepackage{comment}
\usepackage{makecell}
\usepackage{ragged2e}
\usepackage{rotating}
\usepackage{lscape}
\usepackage[table]{xcolor} 
\usepackage{nomencl}
\usepackage{dblfloatfix}
\makenomenclature
\usepackage{wrapfig}
\usepackage{makecell}
\usepackage{balance}
\usepackage[utf8]{inputenc}
\usepackage{ragged2e}
\usepackage{booktabs}% http://ctan.org/pkg/booktabs

\usepackage{pifont}
\usepackage{tabulary}
\usepackage{comment}
\usepackage{lipsum}
\usepackage{array, tabularx, multirow, booktabs, diagbox}
\usepackage{adjustbox}

\newcommand{\cmark}{\ding{51}}%
\newcommand{\xmark}{\ding{55}}%
\begin{document}
\title{Edge-Computing-Enabled Smart Cities: A Comprehensive Survey}
\author{Latif~U.~Khan,~Ibrar~Yaqoob,\IEEEmembership{~Senior~Member,~IEEE,~}Nguyen~H.~Tran,\IEEEmembership{~Senior~Member,~IEEE,~}S.~M.~Ahsan~Kazmi,~Tri~Nguyen~Dang,~Choong~Seon~Hong,\IEEEmembership{~Senior~Member,~IEEE}
 \IEEEcompsocitemizethanks{
 
 \IEEEcompsocthanksitem This work was supported by Institute of Information \& communications Technology Planning \& Evaluation (IITP) grant funded by the Korea government(MSIT) (No.2019-0-01287, Evolvable Deep Learning Model Generation Platform for Edge Computing) and partially supported by the MSIT(Ministry of Science and ICT), Korea, under the Grand Information Technology Research Center support program(IITP-2020-2015-0-00742) supervised by the IITP(Institute for Information \& communications Technology Planning \& Evaluation). *Dr.~C~S~Hong is the corresponding author.
 
 \IEEEcompsocthanksitem L.~U.~Khan,~I.~Yaqoob,~T.~N.~Dang and~C.~S.~Hong are with the Department of Computer Science \& Engineering, Kyung Hee University, Yongin-si 17104, South Korea.
 \IEEEcompsocthanksitem N.~H.~Tran is with School of Computer Science, The University of Sydney, Sydney, NSW 2006, Australia.
\IEEEcompsocthanksitem S.~M.~A.~Kazmi is with the Networks \& Blockchain Lab, Secure System and Network Engineering, Innopolis University, 420500 Tatarstan, Russia, and the Department of Computer Science and Engineering, Kyung Hee University, Yongin-si 17104, South Korea.

%\IEEEcompsocthanksitem Corresponding Author: Choong Seon Hong (cshong@khu.ac.kr)
}
\thanks{
	}}
\IEEEcompsoctitleabstractindextext{%
\justify
\begin{abstract} 
Recent years have disclosed a remarkable proliferation of compute-intensive applications in smart cities. Such applications continuously generate enormous amounts of data which demand strict latency-aware computational processing capabilities. Although edge computing is an appealing technology to compensate for stringent latency related issues, its deployment engenders new challenges. In this survey, we highlight the role of edge computing in realizing the vision of smart cities. First, we analyze the evolution of edge computing paradigms. Subsequently, we critically review the state-of-the-art literature focusing on edge computing applications in smart cities. Later, we categorize and classify the literature by devising a comprehensive and meticulous taxonomy. Furthermore, we identify and discuss key requirements, and enumerate recently reported synergies of edge computing enabled smart cities. Finally, several indispensable open challenges along with their causes and guidelines are discussed, serving as future research directions.    
\end{abstract}
}

\maketitle
\IEEEdisplaynotcompsoctitleabstractindextext
\IEEEpeerreviewmaketitle
\section{Introduction}
\setlength{\parindent}{0.7cm}The unprecedented plethora of miniaturized sensing technologies has shown a remarkable breadth of smart cities vision.
Smart cities enable citizens to experience sustainable, secure, and reliable developments \cite{IoTbasedSmartEnvironments, SC2}. Smart cities are defined formally in numerous ways \cite{smartCitydef1, smartCitydef3, smartCitydef4,smartCitydef5} and thus, there is no absolute definition of the smart cities. Generally, these definitions refer to advancements related to information and communication technologies (ICT) infrastructure, that enable citizens to realize high quality of life via smart services. Smart city services are not limited and can refer to many different processes, resulting in more reliable, secure, sustainable, and innovative cities with unique entrepreneurial opportunities. The key contributors to smart cities are platform developers, research communities, governments, citizens, service providers, and application developers \cite{smartcityStakeholders}. \par  
\setlength{\parindent}{0.7cm}A remarkable increase in city populations is expected in the foreseeable future, which will impose substantial computational complexity regarding advancements in the smart cities. Moreover, this rapid rise in populations will impose further challenges on the city infrastructure due to an exponential increase in data generated by different devices, such as smartphones, global positioning systems, smart sensors, and smart cameras. One way to cope with the challenges of massive scale computing and storage is the use of expensive dedicated computing hardware with sufficient storage capacity. To eliminate the need for dedicated expensive computing hardware, cloud computing was introduced. This is a powerful technology that is intended to enhance the Quality of Experience (QoE) by providing  on-demand storage and processing capabilities in a cost-effective and elastic manner \cite {buyya2009cloud}. The striking features of cloud computing, such as scalability, elasticity, multitenancy, sufficient storage capacity, and resource pooling, have made its adoption possible in different areas \cite{cloudapplicationsreason, cloudfeatures,cloudelasticity}. Recently, cloud computing with virtually unlimited resources has emerged as a promising paradigm to tackle the high computational complexity associated with smart cities \cite{cloudlimits}. However, its inherent limitations of high latency, non-context-aware behavior, and no support for mobility pose serious limitations on its use in real-time smart environments. Apart from these downsides, cloud computing suffers from processing time inefficiency due to the large overhead of smart city device data. On the other hand, edge computing extends the cloud computing resources to the network edge and offers context-awareness, low latency, mobility support, scalability, to name a few. Hence, to address the limitations of cloud computing for enabling real-time smart city environments, edge computing is a viable solution \cite{bringingedge, edge_computing_intro, edge_smartccc,8629866}. \par
\begin{figure*}[!t]
	\centering
	\captionsetup{justification=centering}
	\includegraphics[width=15cm, height=11cm]{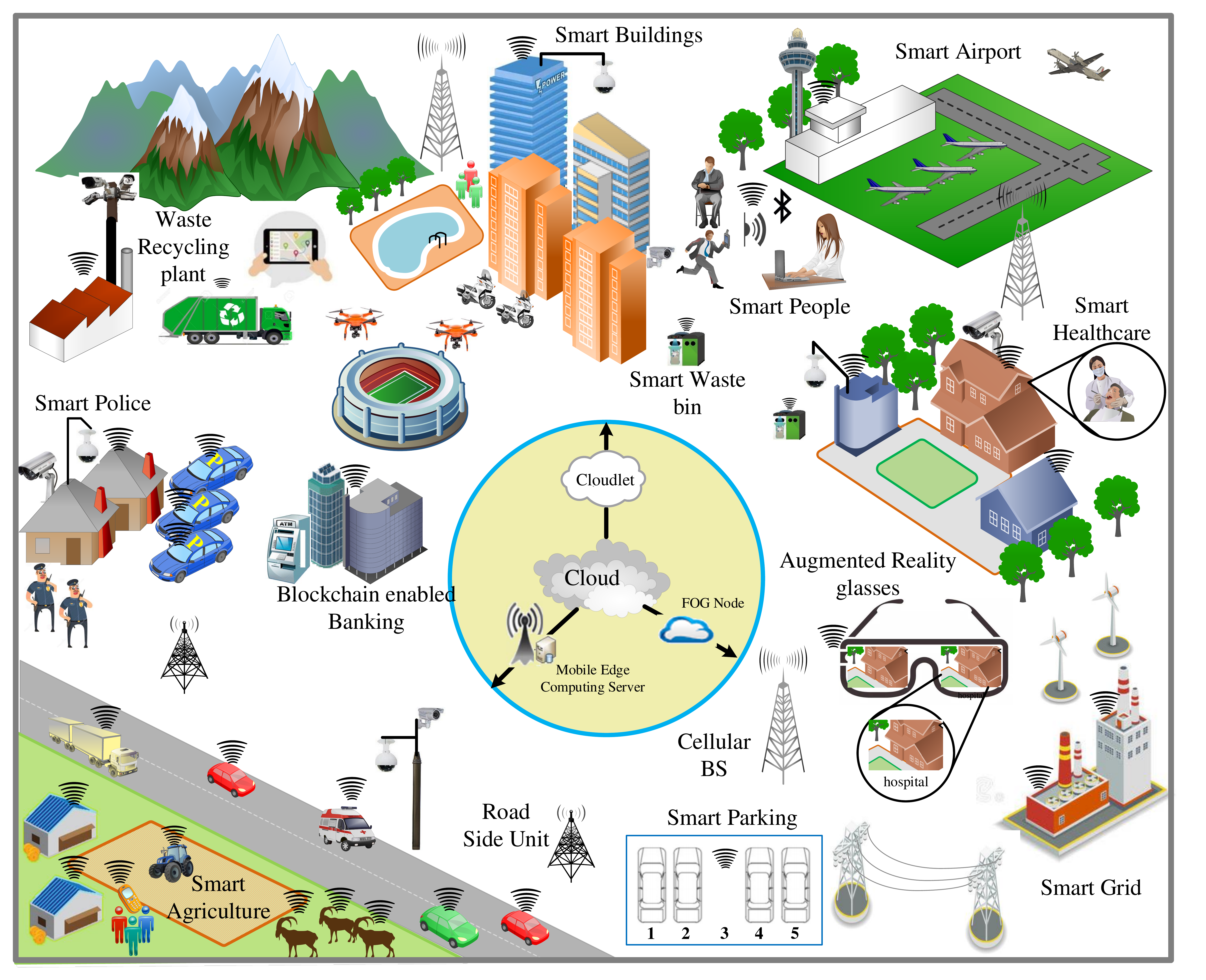}
	\caption{An overview of edge computing enabled smart city.}
	\label{fig:smartcityoverview}
\end{figure*}
\setlength{\parindent}{0.7cm}Figure~\ref{fig:smartcityoverview} elucidates a high-level overview of Internet of Things (IoT) based smart city environments enabled by edge computing. These IoT based smart environments leverage connected smart IoT devices and sensors to improved life standards for citizens. Apart from that, IoT can be defined in different ways because it is associated with a variety of technologies and concepts in literature \cite{IoTdefinition1,IoTdefinition2,IoTdefinition3}. Mostly, the definitions of IoT are inspired from research projects \cite{IoTdefinition4}. In \cite{IoTdefinition1}, the authors found the common and frequently occurring features among different IoT definitions, which include:

\begin {itemize}
\item Existence of a global network infrastructure that enables a unique addressing scheme, seamless integration, and connectivity among heterogeneous IoT nodes.
\item IoT nodes must be addressable, readable, locatable, recognizable, and autonomous. 
\item Existence of heterogeneous technologies that must be given assiduous attention. 
\item Association of services to objects to emphasize its object dependent nature.  
\item Intelligent interfaces between things and humans.
\end{itemize}
%In \cite{IoTdefinition1}, they 
\setlength{\parindent}{0.7cm}We can consider the above-mentioned features (as emphasized in \cite{IoTdefinition1}) in a platform to constitute IoT. However, it may not always be possible for IoT system to possess all of these features. IoT is broadly classified into Industrial Internet of Things (IIoT) and Consumer Internet of Things (CIoT) in \cite{IoTdefinition2}. The IoT assisted applications for smart cities include smart grids, smart lighting systems, smart transportation, smart health-care, smart homes, augmented reality, and smart agriculture \cite{IoTbasedSmartEnvironments,IoTSmartEnvironments2, IoTSmartEnvironments3,IoTSmartEnvironments4}. Therefore, it is evident from literature that IoT is an integral part of smart cities. The next step is enabling the resource intensive and strict latency IoT based smart city applications. Edge computing provides a promising way of enabling these applications by offering computation and storage resources with low latency.\par
Edge computing is a cutting-edge computing paradigm characterized primarily by its geo-distributed operation, context-awareness, mobility support, and low latency. It migrates computing resources, such as computing power, data, and applications, from the remote cloud to the network edge, and thus enables numerous real-time smart city services. Recently, three different emerging technologies, cloudlet, fog computing, and mobile edge computing, have been used in the literature to bring the striking features of cloud computing to the edge \cite{MECSurvey2}. For example, edge computing reduces the network bandwidth usage by alleviating the transmission of data from end users to the remote cloud \cite{bringingedge}.  \par
\subsection {Research Trend and Statistics}
\label{researchtrend}
\begin{figure}[!t]
	\centering
	\captionsetup{justification=centering}
	\includegraphics[width=8cm, height=6cm]{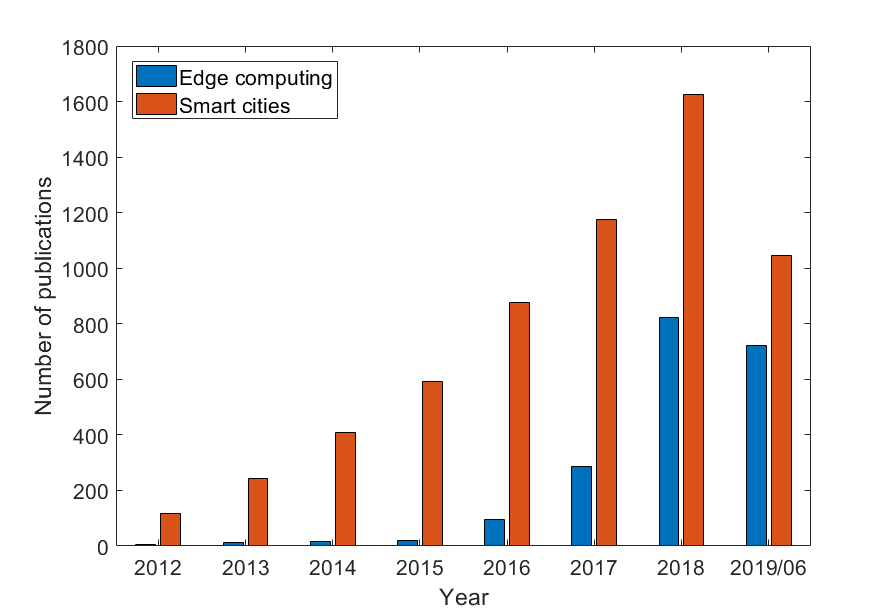}
	\caption{Number of published articles on edge computing and smart cities.}
	\label{fig:researchtrend}
\end{figure}

\setlength{\parindent}{0.7cm}Figure~\ref{fig:researchtrend} illustrates edge computing and smart cities research trends showing the exponential increases in the numbers of publications in both domains. Apart from the research publications trend, according to statistics, the percentages of people living in cities were 29\% and 50\% in the years 1950 and 2008, respectively, which is expected to reach 65\% in 2040 \cite{smartcitiesstatistics1}. Approximately 1.3 million people are moving to cities every week. In 1975, there were three megacities, but currently there are 21 mega cities having a population more than 10 million. Moreover, the number of megacities in 2025 is expected to increase by one city in Africa, two in Latin America, and five in Asia. Furthermore, cities utilize approximately 60\%-80\% of the world energy and generate 50\% of greenhouse gas emissions.\par
The considerable rise in the number of smart cities in the foreseeable future will likely to increase smart cities market share. It has been predicted that the smart cities market share will increase at a Compound Annual Growth Rate (CAGR) of 18.4\%. The smart cities market share will reach USD 717.2 billion by 2023 compared to USD 308 billion in 2018 \cite{statistics2}. The major players of the smart city service providers include Toshiba (Japan), Schneider Electric (France), Siemens (Germany), Hitachi (Japan), Ericsson (Sweden), Huawei (China), Cisco (US), Oracle (US), IBM (US), Microsoft (US), among others. On the other hand, the number of smart IoT devices enabling smart city applications is expected to reach 500 billion by 2030. \par
\setlength{\parindent}{0.7cm}Grand View Research, Inc. estimates the market share of edge computing to reach USD 3.24 billion by 2025 at a CAGR of 41.0\% \cite{statistics3}. Smart health-care is expected to witness the highest CAGR growth in the period from 2017 to 2025. Health-care is expected to exceed the market share of USD 326 million. The potential players in edge computing market are SAP SE, Microsoft, Intel Corporation, International Business Machines (IBM), Huawei Technologies Co. Ltd., Amazon Web Services, Inc. (AWS), Hewlett Packard Enterprise Development LP, General Electric, Cisco Systems, Inc., among others. Edge computing in the market is expected to continually increase. Apart from edge computing market, the rapid rise in IoT based smart cities and edge computing research is evident from the statistics and research trends. These technological advancements in edge computing, IoT, and smart cities have revealed captivating research and development opportunities for manufacturers to enhance their market shares.\par    
\subsection {Motivational Scenarios}
\label{motivationscenario}
\setlength{\parindent}{0.7cm}We provide two motivational scenarios to demonstrate the immense importance of edge computing in developing smart cities. The first scenario deals with reporting of autonomous car accidents, and the second scenario describes the usage of Unmanned Aerial Vehicles (UAVs) with edge computing for smart forest surveillance. \par 
\subsubsection{Autonomous Cars Accident Reporting}
\setlength{\parindent}{0.7cm}A gigantic increase in the number of autonomous cars is expected in the future. There are six different levels of autonomy in autonomous cars ranging from level 0 to level 5 \cite{carAutolevels}. The level 0 represents no automation and level 5 represents full automation, such as operating without driver assistance. Moreover, autonomous cars can handle lane changing, warning signals, collision avoidance, and in-car infotainment, to name a few. To enable real-time analytics for autonomous cars, the use of edge computing enabled Road Side Units (RSUs) is a viable solution. For instance, consider autonomous car accidents, where timely reporting of accidents is required for instant first aid. The key stakeholders of the emergency response system are medical services, smart police, road operators, and edge enabled public safety answering points \cite{carEmergency}. When an autonomous car crash occurs, time is the most critical parameter. The emergency response time must be short to minimize any damage. Reporting such accidents can be performed either by a person or by the edge enabled public answer safety points. Those involved in the crash might suffer from severe injuries and cannot inform authorities about the accident. Therefore, the edge enabled public answer safety points must perform accident evaluation which requires the execution of complex algorithms. To execute such resource intensive and latency sensitive tasks, edge computing is required.\par
\subsubsection{Smart Forest Fire Detection}
\setlength{\parindent}{0.7cm}A smart forest fire detection can substantially reduce damage caused by fires. UAV enabled edge computing offers a method of timely reporting of the forest fire. We can use UAVs assisted by edge computing for surveillance. The UAVs use cameras to continually take images of the forest and requires further processing of these images to determine the presence of fire. One way to determine the presence of fire is to use the remote cloud to perform video images analytics. However, this approach suffers from high latency which is not desirable for immediate action against fires. Processing images at the edge computing enabled UAVs alleviates this latency issue and allows for quick decision making for emergency services. Furthermore, we can also leverage UAVs assisted by edge computing to enable on-demand telecommunication and computing infrastructure to assist in rescue activities \cite{uavenabledEdge, avgeris2019there}.\par
%A smart parking solution enables citizens to experience empty parking slots searching with low-latency.
\subsubsection{Smart Parking System}
Traditional parking systems suffer from challenges of inefficient parking space management and high computational time in finding empty parking slots. To cope with these limitations of the traditional parking systems, one can use smart parking systems which offer instant computation of empty parking slots and efficient parking space management via different schemes based on machine learning \cite {edge_smart_parking1}. The smart parking can be enabled via existing surveillance cameras and image detection systems. These image detection systems use artificial intelligence (AI) schemes (i.e., Convolutional Neural Network) to find the empty parking slots. Therefore, enabling of empty parking slots detection with low-latency requires on-demand computational resources. Edge computing seems to be a promising solution for enabling the smart parking system through on-demand computation resources \cite{8532361,tandon2019optimizing}. \par   
\begin{table*}[]
	\rowcolors{2}{gray!25}{white}
	\caption {Summary of existing surveys and tutorials with their primary focus} \label{tab:surveyssummaries} 
	\begin{center}
		\begin{tabular}{p{2.6 cm}p{1.5cm}p{1.5cm}p{1.5 cm}p{1.5 cm}p{2.5 cm}p{2.5cm}}
			\toprule 
			\textbf{Reference}  & \textbf{Smart city} & \textbf{Internet of Things}& \textbf{Mobile edge computing} & \textbf{Cloudlet} & \textbf{Fog computing} & \textbf{Edge computing evolution}  \\ \midrule
			Bilal \textit{et al.}~,~\cite{kashifbilalsurvey}& \xmark  & \cmark  & \cmark& \cmark & \cmark& \xmark  \\ \midrule
			Shaukat \textit{et al.}~,~\cite{cloudletsurvey1} & \xmark & \xmark  & \xmark & \cmark & \xmark & \xmark  \\ \midrule
			\textit{Ai et al.},\cite{edgetypes} & \xmark & \cmark & \cmark & \cmark & \cmark & \xmark  \\ \hline
			\textit{Mouradian et al.}~,~\cite{fogsurvey1}& \xmark  & \xmark  & \xmark & \xmark & \cmark & \xmark  \\ \midrule
			\textit{Hu et al.}~,~\cite{fogsurvey2}& \xmark  & \cmark  & \xmark & \xmark & \cmark & \xmark  \\ \hline
			\textit{Taleb et al.}~,~\cite{MECsurvey1}& \xmark & \xmark & \cmark & \xmark & \xmark & \xmark  \\ \midrule
			\textit{Abbas et al.}~,~\cite{MECSurvey2}& \xmark  & \cmark  & \cmark & \xmark & \xmark & \xmark  \\ \midrule
			\textit{Roman et al.}~,~\cite{MECSurvey3}& \xmark  & \xmark  & \cmark & \cmark & \cmark & \xmark  \\ \midrule
			\textit{Mao et al.}~,~\cite{MECSurvey4}& \xmark  &  \xmark  & \cmark & \xmark & \xmark & \xmark  \\ \midrule
			\textit{Khan et al.}~,~\cite{edgeSurveyFGCS}& \xmark  & \xmark   &\cmark & \cmark  & \cmark & \xmark  \\ \midrule
			\textit{Hassan et al.}~,~\cite{hassan2018role}& \xmark  & \cmark   &\cmark & \cmark  & \cmark & \xmark  \\ \midrule
			\textit{Yu et al.}~,~\cite{EdgesurveyinIoT}& \xmark  & \cmark   & \cmark & \cmark & \cmark & \xmark  \\ \midrule
			\textit{Gharaibeh et al.}~,~\cite{SmartcityICSTSurvey1}& \cmark  & \xmark  & \xmark  & \xmark & \xmark & \xmark  \\\midrule
			\textit{Lim et al.}~,~\cite{SmartcityICSTSurvey3}& \cmark  & \cmark  & \xmark & \xmark & \xmark & \xmark \\ \midrule
			\textit{Alavi et al.}~,~\cite{SmartcityICSTSurvey4}& \cmark  & \cmark   & \xmark & \xmark &\xmark  & \xmark   \\ \midrule
			\textit{Silva et al.}~,~\cite{SmartcityICSTSurvey5}& \cmark  &  \cmark  &  \xmark &  \xmark & \xmark & \xmark \\ \midrule
			Our Survey & \cmark  & \cmark   &\cmark  &\cmark  &\cmark  &\cmark    \\ 
			\bottomrule 
		\end{tabular}
	\end{center}
\end{table*}
\subsubsection{Smart Home}
A smart home offers enhanced security, smart meters, and smart control \cite{sitton2019edge}. Enabling smart homes with these features requires a multi-layer system with the ability to make a decision regarding home automation using real-time and historical data. Such types of multi-layer systems use different smart IoT devices along with AI techniques to provide various automatic operations at the home. Numerous AI-based smart home appliances are smart washing machines, smart refrigerators, smart speakers, smart TVs, and smart security systems. However, performing smart home tasks based on AI requires instant computational resources at the edge. Therefore, edge computing can be one of the promising solutions to enable smart homes based on AI by offering on-demand computational resources with low-latency.\par

\subsection {Existing Surveys and Tutorials}
\label{existing surveys} 
\setlength{\parindent}{0.7cm}In the literature, numerous surveys and tutorials have been presented to provide considerable insights into edge computing and smart cities \cite{kashifbilalsurvey, cloudletsurvey1, edgetypes, fogsurvey1, fogsurvey2,MECsurvey1, MECSurvey2, MECSurvey3, MECSurvey4,edgeSurveyFGCS, hassan2018role, EdgesurveyinIoT, SmartcityICSTSurvey1,SmartcityICSTSurvey3,SmartcityICSTSurvey4,SmartcityICSTSurvey5}. A summary of recently published surveys along with their primary contribution areas is given in table~\ref{tab:surveyssummaries}. \par

\subsubsection{Edge Computing Surveys}
\setlength{\parindent}{0.7cm}Several surveys pertaining to edge computing have been conducted in \cite{kashifbilalsurvey, cloudletsurvey1,edgetypes, fogsurvey1, fogsurvey2,MECsurvey1, MECSurvey2, MECSurvey3, MECSurvey4,edgeSurveyFGCS, hassan2018role, EdgesurveyinIoT }. In \cite{kashifbilalsurvey}, the authors discussed different edge computing paradigms, such as cloudlets, mobile edge computing, and fog computing. Additionally, they discussed applications and summarized the architectures of state-of-the-art edge computing paradigms. Finally, they outlined open research problems regarding edge computing. A survey of cloudlet architectures, cloudlet solution taxonomy, and challenges pertaining to the deployment of cloudlets in local wireless networks is presented in \cite{cloudletsurvey1}. The taxonomies are devised using three different parameters: augmentation models, cloudlet service models, and architecture. Furthermore, cloudlet deployment requirements are also discussed. In \cite {edgetypes}, Ai \textit{et al.} presented a study on cloudlets, fog computing, and mobile edge computing. Specifically they presented general architecture, standardization efforts, applications, and design principles of the edge computing paradigms. In addition, open research issues are also highlighted in their paper. Mouradian \textit{et al.} in \cite{fogsurvey1} proposed a comprehensive survey on fog computing architecture and algorithms. The authors proposed evaluation criteria, such as heterogeneity management, scalability, mobility, federation, and interoperability. The evaluation criteria are further used to evaluate the literature related to architecture and algorithms for fog computing. Additionally, the architectural and algorithmic open research problems are also outlined. A comprehensive survey proposed in \cite{fogsurvey2} discussed fog computing architecture, applications, key technologies, and research challenges. The authors proposed a hierarchical architecture of fog computing, followed by a discussion of key technologies. Apart from that, they also discussed various smart environments. Finally, open research challenges are also presented. In \cite{MECsurvey1}, Taleb \textit{et al.} presented the evolution of edge computing, use cases of mobile edge computing, key mobile edge computing enablers, reference architecture, deployment challenges, and open research challenges. Abbas \textit{et al.} discussed architecture, emerging smart applications, security and privacy, and open research challenges in mobile edge computing\cite{MECSurvey2}. The authors in \cite{MECSurvey3} provided a survey on security and privacy issues in mobile edge computing. In addition, a detailed overview of edge paradigms concepts, challenges, and security threats are also presented. They identified threat models and discussed guidelines for their solutions. Furthermore, a state-of-the-art security mechanism used in edge computing paradigms is also outlined. The study in \cite{MECSurvey4} presented a comprehensive survey on computation and radio resource management in mobile edge computing. They discussed green mobile edge computing, privacy-aware mobile edge computing, mobility management for mobile edge computing, and cache enabled mobile edge computing. Along with this, open research challenges are also discussed. In \cite{edgeSurveyFGCS}, Khan \textit{et al.} presented survey on edge computing. They have discussed different edge computing paradigms, such as cloudlets, mobile edge computing, and fog computing. key characteristics of edge computing are presented. Additionally, fundamental requirements and open research challenges for realizing edge computing are presented. In another study in \cite{hassan2018role}, the role of edge computing in IoT has been discussed. Taxonomy has been devised based on edge computing enabled IoT. Furthermore, numerous smart applications, key requirements, and few open research challenges are presented. Yu \textit{et al.} in \cite{EdgesurveyinIoT} conducted a comprehensive survey on the role of edge computing in IoT networks. They classified the edge computing based on architecture into the far-end, near-end, and front-end. Furthermore, they discussed the advantages of edge computing-based IoT infrastructure and reported open research challenges. \par
\subsubsection{Smart Cities Surveys}
\setlength{\parindent}{0.7cm}In the past, several surveys have been conducted on smart cities \cite{SmartcityICSTSurvey1,SmartcityICSTSurvey3, SmartcityICSTSurvey4, SmartcityICSTSurvey5}. Gharaibeh \textit{et al.} in \cite{SmartcityICSTSurvey1} surveyed data management techniques used for ensuring reuse-ability, granularity, interoperability, and consistency of smart IoT device data. Additionally, the authors identified techniques used for security and privacy in smart cities. Other than that, emerging technologies enabling smart cities are also discussed. In \cite{SmartcityICSTSurvey3}, the authors discussed the use of big data along with different projects that enable the development of smart cities. The authors have classified the use of urban big data into four kind of reference models. Furthermore, the challenges of transforming smart city data into information are also presented. The study \cite{SmartcityICSTSurvey4} discussed IoT based smart city applications. The authors provided an overview of current IoT based projects. Additionally, a prototype for a real-time monitoring system using IoT for smart cities has been presented. An overview of smart cities characteristics, architecture, developments, and challenges has been presented in \cite{SmartcityICSTSurvey5}.   
\subsubsection{Our Survey}
\setlength{\parindent}{0.7cm}To the best of our knowledge, we are the first to jointly consider (as can be seen in table~\ref{tab:surveyssummaries}) smart cities, IoT, and edge computing. Previously published surveys and tutorials reviewed smart cities with IoT \cite{SmartcityICSTSurvey3,SmartcityICSTSurvey4,SmartcityICSTSurvey5}, whereas \cite{SmartcityICSTSurvey1} considered smart city only. However, \cite{kashifbilalsurvey,cloudletsurvey1,edgetypes,fogsurvey1,fogsurvey2, MECsurvey1,MECSurvey2,MECSurvey3,MECSurvey4,EdgesurveyinIoT} reviewed either edge computing alone or edge computing with IoT. Unlike the existing surveys, we comprehensively present premier smart city environments enabled by edge computing, the evolution of edge computing, taxonomy, core requirements, and open research challenges. 
\begin{figure}[!t]
	\centering
	\captionsetup{justification=centering}
	\includegraphics[width=8cm, height=17cm]{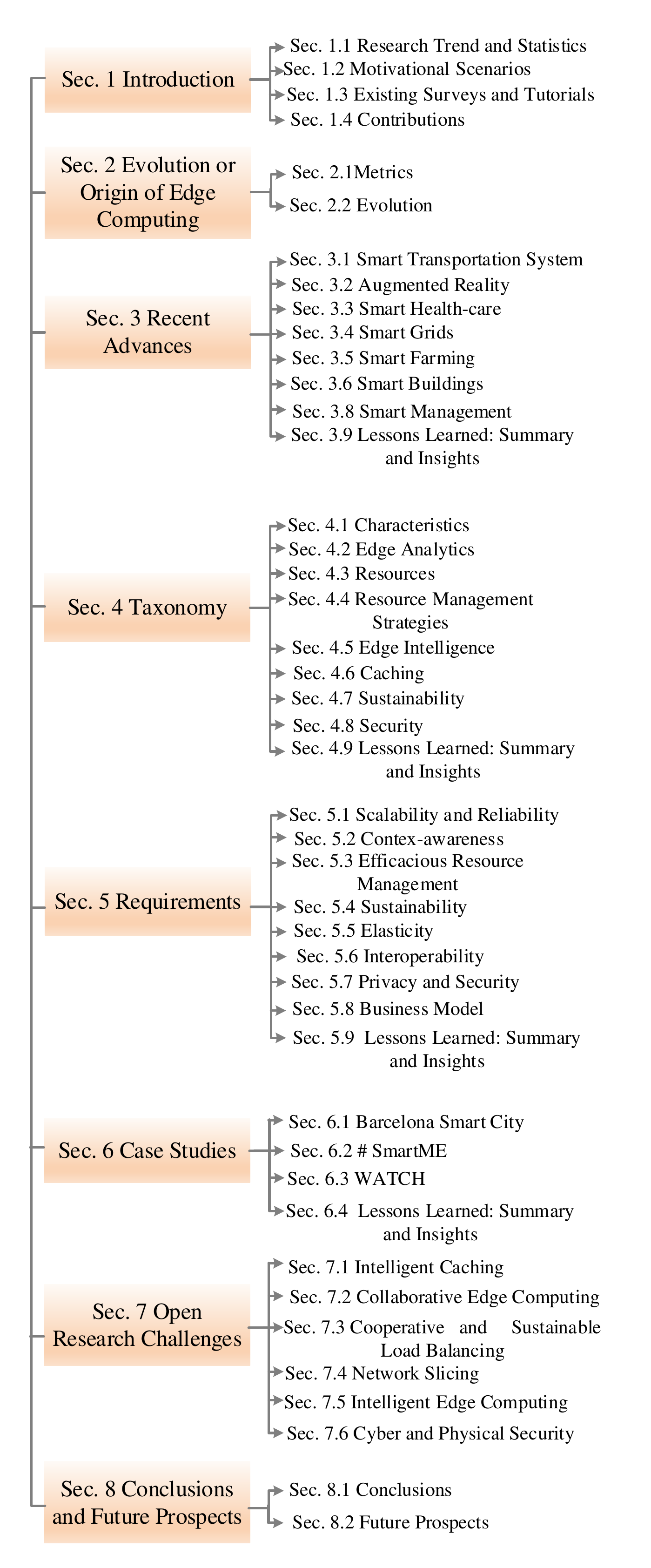}
	\caption{Structure of the survey.}
	\label{fig:strucuture}
\end{figure}
\subsection {Contributions}
\label{contributions} 
\setlength{\parindent}{0.7cm}The key contributions of this survey are as follows:
\begin{itemize}
  \item We present the evolution of edge computing using scalability, flexibility, cost optimization, automation, mobility, and latency, as metrics.
  \item We report, critically analyze and evaluate the premier advances made in edge computing enabled smart cities.
  \item We categorize and classify the literature by devising a comprehensive and meticulous taxonomy using important parameters, such as characteristics, security, edge analytics, resources, edge intelligence, resource management strategies, caching, and sustainability. 
  \item A few indispensable requirements for the design of the smart cities architectures using edge computing are presented. 
  \item Several open research challenges along with their causes and guidelines regarding the implementation of edge computing in smart cities are discussed.   
\end{itemize}

\setlength{\parindent}{0.7cm}The rest of the paper is organized (illustrated in figure~\ref{fig:strucuture}) as follows: Section 2 discusses the evolution of edge computing. Recent advances and literature categorization are given in section 3. Section 4 presents the devised taxonomy based on different parameters. The core requirements in designing edge computing enabled smart cities are given in section 5. The recently reported synergies and case studies along with their key objectives, organizations involved, and deployment year are given in section 6. Section 7 identifies, discusses, and provides guidelines for solutions of the indispensable open research challenges. Finally, the paper is concluded in section 8. 
\section{Evolution or Origin of Edge Computing}
\label{sec:Edgecomputing}
\setlength{\parindent}{0.7cm}In this section, we consider the chronological development and attributes of different computing paradigms, namely, mainframe computing, mini computing, client server computing, desktop cloud computing, Mobile cloud Computing (MCC), and edge computing. Similar to \cite{abolfazli2014rich}, two paradigms, functionalism and structuralism are considered in this paper to extend our findings. Structuralism enables identification of the building blocks and their relationships to provide a clearer understanding of the phenomenon, whereas functionalism provides insights into the behavior and features of a phenomenon through analysis of its role in present and future with functionalities. \par
\begin{figure*}[!t]
	\centering
	\captionsetup{justification=centering}
	\includegraphics[width=18cm, height=11cm]{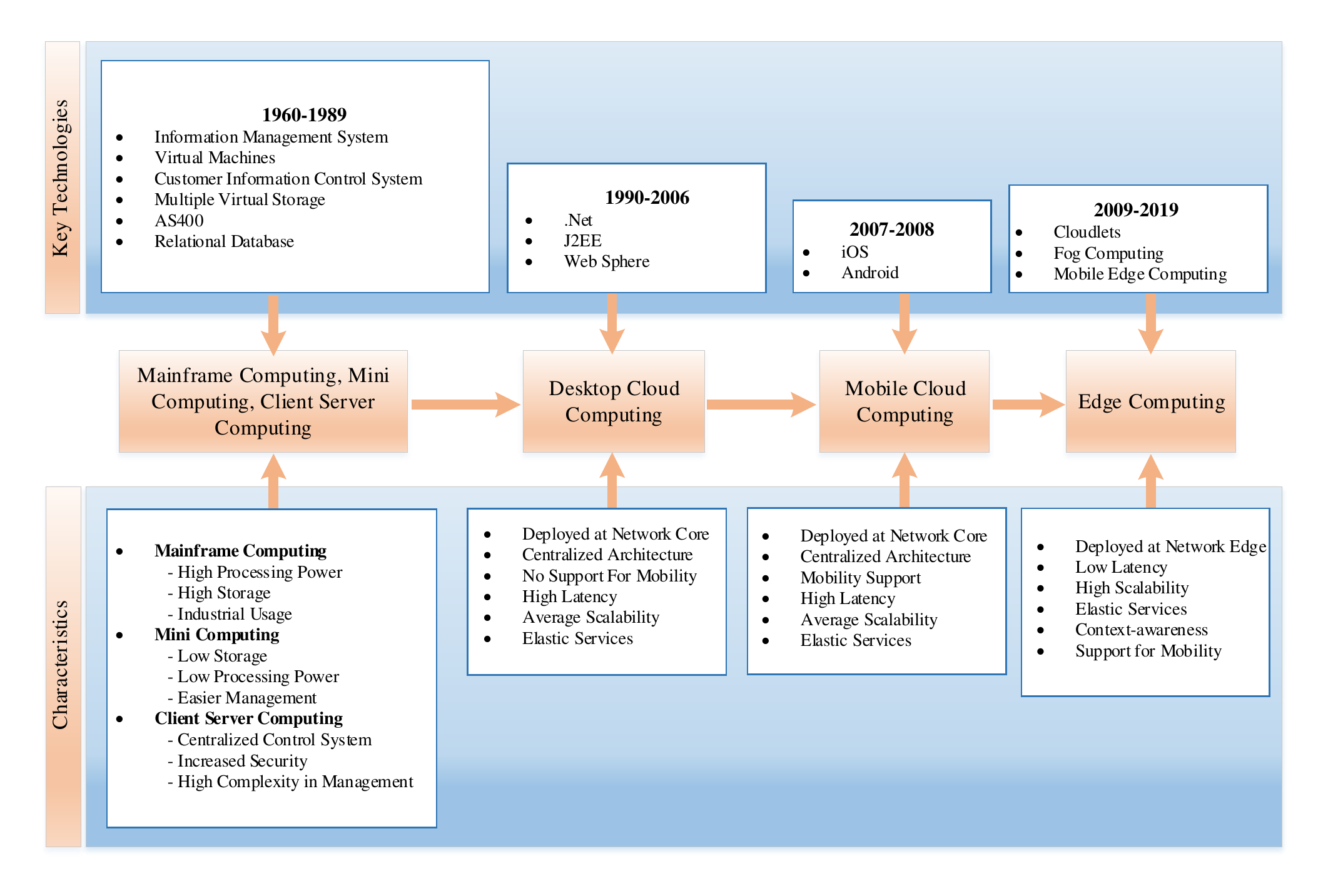}
	\caption{Evolution of edge computing.}
	\label{fig:genesis}
\end{figure*} 
\subsection{Metrics}
\label{Metrics}
\setlength{\parindent}{0.7cm}We consider six metrics, namely, scalability, flexibility, cost optimization, automation, latency, and mobility support for evaluation of different computing paradigms:
\begin{itemize}
  \item \textbf{Scalability:} This parameter reflects the ability of a system to expand its services in response to increasing demands of the users.   
  \item \textbf{Flexibility:} An ability of the system to provide computing infrastructure in an elastic way. 
  \item \textbf{Cost Optimization:} This refers to the ability of a system to provide users on-demand computing resources for optimizing the overall cost. In addition, it also refer to the cost associated with hardware and software in the acquisition of the computing resources.   
  \item \textbf{Automation:} The ability to perform cloud updates without any intervention of the end users.  
    \item \textbf{Latency:} This parameter is used to measure the total execution time of smart city applications.
    \item \textbf{Mobility Support:} It aims to provide an ability for enabling seamless execution of IoT-based applications. 
\end{itemize}
\subsection{Evolution}
\label{Evolution}
\setlength{\parindent}{0.7cm}Figure~\ref{fig:genesis} presents the gradual evolution of edge computing proceeds by other computing technologies, such as mini computing, mainframe computing, client-server computing, desktop cloud computing, and Mobile cloud Computing (MCC). Further details are provided below.   
\subsubsection{Mainframe, Mini, and Client-Server Computing}
\setlength{\parindent}{0.7cm}In the early era of computing, mainframe computers that include ENIAC, Markl, BINAC, Whirlwind, UNIVAC, IBM 701, and IBM 360, were primarily used to enable bulk data processing. In contrast, minicomputers (developed by IBM) have a smaller size than mainframe computer and are mainly used as mid-size servers to process scientific applications. Different from mainframe computing and mini computing, client-server computing is based on providing resources to clients on request. Xerox PARC was the first client-server computing model introduced in the 1970s.\par
\subsubsection{Desktop Cloud Computing}
\setlength{\parindent}{0.7cm}The concept of cloud computing dates back to 1963, when the Defense Advanced Research Projects Agency (DARPA) provided \$2 million to MIT for project Project on Mathematics and Computation (MAC) \cite{cloudstats1}. The purpose of project MAC was to establish a technology that enables simultaneous use of a computer by multiple people. In 1969, the Advanced Research Projects Agency Network (ARPANET) was developed by J. C. R. Licklider. The vision “\textit{Intergalactic Computer Network}" (also known as the Internet) was promoted by Licklider to enables connectivity between all humans on the earth. Later in 1997, Ramnath Chellappa of Emory University devised the definition of cloud computing as a "\textit{computing paradigm, where the boundaries of computing will be determined by economic rationale, rather than technical limits alone}". \par
\setlength{\parindent}{0.7cm}The paradigm of cloud computing began gaining more interest when the different organizations started exploiting it to provide services to the end-users. In 1999, Salesforce successful implemented cloud computing to provide software services using the Internet. The end-users were given the ability to download an application on demand in a cost-effective way from a remote location \cite{evolution_edge}. Amazon in 2002 offered its retail services through the Web. In 2006, Amazon started providing web-based services to clients on-demand. Google also started providing Google doc services in the same year. Following this, IBM in 2011 offered IBM SmartCloud and Apple introduced iCloud to enable storage of personal information, such as videos, music, photos, among others. In 2012, Oracle started offering Oracle Cloud which enabled three categories of services: SaaS (Software-as-a-Service), PaaS (Platform-as-a-Service), and IaaS (Infrastructure-as-a-Service).\par 
\subsubsection{Mobile Cloud Computing}
\setlength{\parindent}{0.7cm}MCC was introduced to enable mobile computing with the ability to leverage cloud computing. Mobile devices have inherent limitations of frequent disconnections, resource scarcity, and energy constraints. Therefore, to execute high complexity tasks, it is preferable to execute a resource-intensive task on a remote cloud. MCC has been defined in three ways in \cite{mobilecloudsurvey}:
\begin{itemize}
  \item MCC refers to the running of mobile device applications (e.g. Google's Gmail) on a remote server. 
  \item MCC deals with the use of mobile devices and other stationary devices resources as a cloud resource. Mobile devices form a peer-to-peer network and allow use of their resources by other devices. 
  \item MCC refers to a cloudlet which is a micro data center consisting of several multi-core computers and which provides computation resources to nearby mobile devices \cite{satyanarayanan2009case}.  
\end{itemize}
\setlength{\parindent}{0.7cm}In this paper, we use the term MCC to mean running of client applications of mobile devices on remote servers, such as Google servers and Amazon servers. On other hand, various operating systems, such as iPhone OS (IoS) and android, which were introduced in 2007 and 2008, respectively, provide mobile devices with more capabilities by using apps.  
\subsubsection{Edge Computing}
\setlength{\parindent}{0.7cm}Edge computing refers to practices that move computing and storage resources to the network edge. In the literature, three different practices, such as cloudlets, fog computing, and mobile edge computing have been used, that extend cloud computing to the network edge. All of these three terms used in the literature refer to edge computing \cite{edgetypes, satyanarayananedgeemergence, MECSurvey2}. Edge computing is distributed architecture that enables real-time data processing and acceleration of data-streams with low latency. Along with this, edge computing also minimizes the Internet bandwidth usage, which further reduces network congestion and delay. \par
\setlength{\parindent}{0.7cm}The concept of cloudlets was introduced in \cite{satyanarayanan2009case} by Satyanarayanan \textit{et al}. to place a small scale data center near the network edge to enable operation of resource-intensive applications with low latency. A cloudlet represents a middle layer in an architecture having three tiers: a centralized cloud, edge cloud platform (cloudlet), and end device \cite{MECsurvey1}. CISCO coined the term \textit{fog computing}, which refers to an architecture that allows the use of edge devices to provide computation services near the network devices \cite{FOG}. Similar to cloudlets, fog nodes (edge devices) represent the middle tier of a three-tier hierarchy that consists of cloud, fog nodes, and end devices. Fog nodes cover a wide range of devices, such as routers, wireless access points, and data centers \cite{fognodes}. \par
\setlength{\parindent}{0.7cm}The term mobile edge computing was introduced by European Telecommunications Standards Institute (ETSI) to enable the placement of computing resources at the radio access network and core network of mobile telecommunication systems \cite{MEC,MECdef}. The mobile edge computing servers can be placed at different positions in the radio access network such as a macrocell BS \cite{MEC2, MEC3}, assigned to a group of BSs \cite{MEC4}, or in a hierarchical fashion at different levels \cite{MEC5}. Consider table~\ref{tab:edgecomparison}, which lists the differences between cloud computing, mobile edge computing, cloudlets, and fog computing \cite{ullah2018information}. Mobile edge computing differs from cloudlets and fog computing in its primary scope, which is telecommunication networks. In terms of context-awareness, mobile edge computing has the highest value while cloud computing has no context-awareness. Fog computing and cloudlets have medium and low context-awareness. Other than context-awareness, mobility is supported by all three categories of edge computing, whereas cloud computing has no or limited mobility support. All of the four computing paradigms such as cloud computing, cloudlet, fog computing, and mobile edge computing offer scalability and reliability \cite{MECSurvey3}. Both cloudlets and fog computing have low deployment cost compared to cloud computing due to low-cost availability of wireless access points and routers \cite{ullah2018information}.

% % % % % % % % % % % % % % % % % % % % % % % % % %
\begin{table*}
\caption {Edge computing paradigms comparison} 
\label{tab:edgecomparison} 
  \centering
  \begin{tabular}{lllllll}
    \toprule
   &\textbf{ Cloud computing }& \textbf{Cloudlets} & \textbf{Fog computing} & \textbf{Mobile edge computing} \\
\hline

\textbf{Context-awareness} & No & Low & Medium & High \\ \midrule
\textbf{Geo-distribution} & Centralized & Distributed & Distributed & Distributed \\\midrule
\textbf{Latency} & High & Low & Low & Low \\\midrule
\textbf{Mobility support} & No/Limited & Yes & Yes & Yes \\\midrule
%\textbf{Primary scope} & Telecommunication \& IoT   &  IoT & IoT & Telecommunications \\\midrule
\textbf{Distance} & Multi hop & Single hop & Single hop/ Multi hop & Single hop \\\midrule
\textbf{Scalability} & Yes & Yes & Yes & Yes \\ \midrule
\textbf{Flexibility} & Yes & Yes & Yes & Yes \\ \midrule
\textbf{Deployment cost} & High & Low & Low & High \\ 
  \bottomrule
  \end{tabular}
\end{table*}
\section{Recent Advances}
\label{sec:stateoftheart}
\setlength{\parindent}{0.7cm}This section critically analyzes the dominant trends toward the edge computing enabled smart cities. Apart from providing insight into the architectural and algorithmic aspects, we perform a rigorous evaluation of the literature using various assessment parameters. These assessment parameters are primarily derived from the literature of recent advances in edge computing enabled smart cities \cite{smarttransportation2,huang2017exploring, smartTransportation3, augmentedreality3, taleb2017mobile, augmented3, augmented4, muhammad2018edge, healthcare2, healthcare3, kumar2016vehicular, smartgrid2, smartgrid3, ferrandez2018precision, smart_agriculture2, smartfarming3, vallati2016mobile, smarthome, smartgaming, smartManagement1}. We consider five different assessment parameters, namely, context-awareness, scalability, sustainability, caching, and security for evaluation of the recent advances: 
\begin{itemize}
\item \textbf{Context-awareness:} An ability of the system to obtain information pertaining to the nodes locations and surrounding environment. Context-awareness offers several advantages including the addition of more meaning to smart IoT device data, M2M communication facilitation, emergency management assistance, and automatic execution of smart city services, to name a few. \par
\item \textbf{Sustainability:} This deals with energy efficient design and use of renewable energy sources. In addition, sustainability refers to energy harvesting, which takes energy from environmental sources and radio frequency sources and stores it for further use. The primary objective of sustainability is to reduce the carbon footprint. A significant portion (more than 80\%) of energy is generated using brown sources \cite{sustainabilityEDGE}. Therefore, sustainability is an important aspect that must be considered in the design of edge computing enabled infrastructure. Additionally, this metric offers an increase in profit, reduction in carbon footprint, and hardware reliability \cite{buyyasustainability}.  \par 
\item \textbf{Scalability:} The ability of a system to enable elastic services as per user demands without losing QoS, resulting in a cost-efficient operation. \par
\item \textbf{Caching:} The temporary storing of popular content at different locations in a network to enable access with low latency. Caching also reduces congestion in a network by avoiding repeated traffic flow.Note that caching deals with the storage of popular content and edge computing provides computation resources, but these two concepts can be leveraged simultaneously in smart cities to enable a variety of smart applications.\par
\item \textbf{Security:} This refers to a devices physical security and cyber security. More specifically, cyber security deals with the protection of the network, computing infrastructure, and data from attacks.
\end{itemize}
\setlength{\parindent}{0.7cm}Table~\ref{tab:performance_parameters} summarizes explanation, key advantages, and dimensions of the parameters to evaluate recent advances. Every parameter has a set of dimensions. For instance, sustainability has architectural, algorithmic, communication, renewable energy, software, and hardware dimensions. All of these dimensions are key drivers of sustainability in edge enabled smart cities. We use a check mark (\cmark) for a parameter if has one or more dimensions are considered and cross mark (\xmark) if its no dimension that is considered. Table~\ref{tab:recent_advances_evaluation1}-\ref{tab:recent_advances_evaluation2} highlight key contributions and categorize the recent advances according to different applications. Moreover, these tables present an evaluation of the literature. 
% % % % % % % % % % % % % % % % % % % % % % % % % %
\begin{table*}[]
\caption {Evaluation parameters: explanation, dimension, and advantages} \label{tab:performance_parameters} 
\begin{center}
\begin{tabular}{lp{4cm}p{4cm}p{4cm}}
\toprule
\textbf{Parameter }    & \textbf{Explanation}  & \textbf{Dimension} & \textbf{Advantage} \\ \midrule
\textbf{$P_1$: }\textbf{Context-awareness}  & It deals with locations of smart devices and their surrounding environment. & \begin{itemize} \item Context-aware architecture    \item  Context-aware algorithm \end{itemize} & \begin{itemize} \item Helps in accurate decision-making \item Efficient emergency management assistance \item  Enables automated smart services\end{itemize}\\ \midrule
\textbf{$P_2$: }\textbf{Scalability}   & It refers to an ability of a system to operate effectively with an increase in demand by adding more resources elastically.  &   \begin{itemize}  \item Scalable architecture  \item Scalable hardware \item Scalable Software \item Scalable algorithm   \end{itemize} & \begin{itemize} \item Cost optimization \item QoS maintenance  \item Instant service availability \end{itemize} \\ \midrule
\textbf{$P_3$: }\textbf{Sustainability}  & It deals with energy efficient design and use of renewable energy sources to minimize the carbon footprint. & \begin{itemize} \item Sustainable architecture \item Sustainable Software \item Sustainable hardware   \item Sustainable algorithm \item Sustainable communication \item Renewable energy sources \end{itemize}  & \begin{itemize} \item Carbon footprint reduction \item Profitable infrastructure  \item Hardware reliability  \end{itemize}  \\ \midrule
\textbf{$P_4$: }\textbf{Caching}  & It allows temporary storage of popular contents at different locations in a network. & \begin{itemize}\item Caching architecture \item Caching algorithm \item Caching hardware \item Caching software  \end{itemize} &  \begin{itemize} \item  Instant access to popular contents  \item  Back-haul traffic reduction  \end{itemize}\\ \midrule
\textbf{$P_5$: }\textbf{Security}  &    It deals with physical and cyber security of smart devices and servers involve in edge computing. & \begin{itemize}  \item Security algorithm \item Security architecture \end{itemize}  &  \begin{itemize} \item Continuous service availability  \item Users privacy preservation \end{itemize}  \\ 
\bottomrule
\end{tabular}
\end{center}
\end{table*}

%%%%%%%%%%%%%%%%%%%%%%%%%%%%%%%%%%%%%%%%%%%%%%%%%%%%%%%%%%%%%%%%%%%%%%%%

\subsection{Smart Transportation System}
\setlength{\parindent}{0.7cm}A smart transportation system is envisioned to enable innovative traffic management schemes, advanced transport modes, autonomous driving, and in-car infotainment services. \par
\subsubsection{Smart Transportation Applications}
\setlength{\parindent}{0.7cm}Giang \textit{et al.} in \cite{smarttransportation2} discussed the development of smart transportation applications leveraging fog computing. They have discussed the requirements, state-of-the-art, and research challenges regarding implementation of the fog computing enabled smart transportation applications. Although the authors discussed the requirements of scalability and context-awareness in fog computing assisted vehicular networks they did not consider service migration \cite{taleb2017mobile} and security \cite {vechicularnetworksSecurity}, which have significant importance in vehicular networks. The vehicle might change connectivity from one roadside unit to another during video streaming. Therefore, service migration must be considered in the design of fog assisted application for vehicles. Similarly, the security aspect must be addressed to avoid any accidents and physical damage.  \par
\subsubsection{5G-enabled Software Defined Vehicular Networks} 
\setlength{\parindent}{0.7cm}In \cite{huang2017exploring}, the authors proposed a 5G-enabled software defined vehicular network (5G-SDVN) architecture, that consists of three planes, such as data plane for the transmission of data, social plane for Vehicular Neighbor Groups (VNGs) networking, and control plane with mobile edge computing. The key contributions of the authors include the discovery of VNGs using real data set and the integration of software-defined networking (SDN) with mobile edge computing to enable flexible architecture. In addition, the proposed architecture provides the benefits of efficient data sharing, ubiquitous mobile interaction, and reliable resource cooperation. Apart from these benefits, SDN offers separation of the control logic from the underlying infrastructure, which provides more freedom for adding functionality (i.e.intelligence) to a network regardless of change in hardware. Although, the proposed architecture offers significant advantages, but it does not consider the case of active service migration as in \cite{smarttransportation2}.
\subsubsection{Cost Optimized Edge Enabled Transportation}
\setlength{\parindent}{0.7cm}The study \cite{smartTransportation3} considered a smart city scenario where vehicles ran applications that take data from the environment and send them to edge computing servers through roadside units (RSUs) to enable further processing and analysis. The edge computing servers are co-located with the RSUs, which are further connected to the cloud through the Internet. In a vehicular network, maintaining seamless connectivity between the vehicles and RSUs is difficult. On the other hand, the computational capacities of the edge servers also have limitations. Therefore, it is imperative to deploy the RSUs to enable effective resource management and seamless connectivity. The objective of the paper is to minimize the network deployment cost while fulfilling the minimum required QoS for vehicular applications. The problem is formulated as a mixed integer linear programming problem, whose goal is to jointly minimize the total cost associated with the deployment of RSUs, the cost associated with assigned power levels to each RSU, and the cost associated of the distances of RSUs from the cells it serves (the target area is divided into cells). A single RSU can serve more than one cell. Therefore, only the cells adjacent to an RSU cell must be considered as opposed to distant cells, which increase the cost.  

\subsubsection{Lessons Learned: Summary and Insights}
Herein, we summarized various applications of edge computing in enabling smart transportation, and discussed lessons learned, insights, and future enhancements:
\begin{itemize}
    \item Security is one of the important parameters that must be considered in the design of edge computing enabled smart transportation systems. For example, if a hacker accesses the vehicle and alters the lane change information, this might result in accidents, specifically in autonomous cars.
    \item The challenge of active service migration which arises due to the mobility of vehicles must be tackled. For instance, moving infotainment enabled smart cars must have seamless connectivity with the edge server-based RSU during contents downloading \cite{kazmiinfortainment}. A set of contents stored at a RSU edge computing server must be provided seamlessly to the requesting passenger in a car. The passenger in a moving car might lose its connectivity with the existing RSU and moved to a new RSU range during the content downloading phase. This challenge can be handled using effective service migration schemes \cite{taleb2017mobile}. For example, one way is to migrate the content to the new RSU for serving the user. Another way is the use of content placement at the shared memory placed in close vicinity to end users at a common location with fast connectivity to different RSUs. A requesting end user can access the contents from the shared memory through its associated RSU. 
    \item The set of edge computing server-based RSUs must be deployed in a fashion that simultaneously minimizes the cost associated with transmission power, number of RSUs, and interference level for improving QoS.
\end{itemize}

\subsection{Augmented Reality}
\setlength{\parindent}{0.7cm}Augmented Reality  (AR) enabled smart cities provide users with smart Industrial Augmented Reality, smart tourism, smart remote live support, smart web-based augmented reality, as we discuss in the following subsections.\par %
\subsubsection{Smart Industrial Augmented Reality}
\setlength{\parindent}{0.7cm}The authors in \cite{augmentedreality3} proposed an industrial ARarchitecture based on fog computing, named, Navantia’s industrial AR(IAR). The use of IAR in smart industries leads to enhancing productivity. The proposed IAR architecture consists of three layers, such as IAR layer, edge layer, and cloud layer. The IAR layer includes ARdevices that use WiFi along with other wireless technologies for connectivity with the edge layer. The IAR layer devices interact with the edge layer through local edge layer gateways that provide low-latency services, such as data caching and sensor fusion. The authors considered both scalability and caching in their architecture. However, they did not consider security which is required for safe operation. \par
%For example, an unauthorized user can access the system and interrupt proper operation of the system, which can lead to serious consequences.\par
\subsubsection{Smart Tourism}
\setlength{\parindent}{0.7cm}In  \cite{taleb2017mobile}, Taleb \textit{et al.} envisioned mobile edge computing enabled architecture for high definition video streaming and considered two use cases of smart tourism. One case is interactive glasses to obtain information about the visited site, whereas the other case involves tourists taking videos, commenting on these videos, and sending them to the edge computing server. Although migration of such a service from one edge server to another is considered in their work, but they did not consider resource management for multiple users, which is more realistic than a single user scenario. Generally, tourists visit historical places in groups (social communities), therefore device-to-device communication with social awareness must also be considered for performance improvement. The user may request content from another user that already has it instead of sending a request from the access point. This will improve the transmission rate of the requesting user and reduce congestion of the access point. \par 
\subsubsection{Smart Remote Live Support}
\setlength{\parindent}{0.7cm}Edge computing enabled AR has been considered in \cite{augmented3} to enable smart management of industrial tasks. The authors considered a scenario of remote live support that uses guidance from the remote expert through AR and video transmission to tackle complex faults in a machine. The main parts of the proposed architecture include a client device, edge server, and a remote device. The job of the client device with an integrated camera and running an operating system is to grab video frames for further transport to the edge server. The edge server process all the computationally expensive operations. Additionally, the edge server forwards the video stream to a remote expert. The remote expert has the privilege of pausing the video for addition of hints in it. The video contents are sent back to the edge server, which are further sent back to the client after rendering annotations into the camera feed. The author considered compression of the video data prior to transmission. This adds further delay in addition to transmission delay, which is not desirable. Apart from that delay, the processing delay must not exceed the maximum delay limit. The better option is to use communication technologies with higher data rates instead of compression.\par
\subsubsection{Smart Web-based Augmented Reality}
\setlength{\parindent}{0.7cm}In \cite{augmented4},  Qiao \textit{et al.} considered web-based AR enabled by edge computing. The latency sensitive and resource-intensive applications of the web-based AR are hampered by the weak computational efficiency of web browsers. To enable resource intensive and latency sensitive web-based AR applications, web-based AR assisted by edge computing is proposed by the authors. Different from edge computing, cloud computing can be leveraged to assist computation intensive web-based AR applications. However, this approach causes an increase in latency due to the remote location of the cloud. The key contributions of the authors include a proposal of edge computing assisted web-based augmented reality, design details, and deployment challenges.   \par

\subsubsection{Lessons Learned: Summary and Insights}
In this section, we have discussed different smart applications based on augmented reality enabled by edge computing. Several lessons learned and future recommendations derived from this section are as follows:
\begin{itemize}
    \item We derived the use of socially-aware device-to-device communication in edge computing-based AR for smart tourism. A set of users requesting the same content from the edge computing-based BS might result in performance degradation due to the extra overhead of requests. On the other hand, a user can first check the requesting content within its social group before requesting to BS. Downloading the content from the nearby devices significantly reduces latency compared to requesting from the BS. Additionally, the requests overhead at the BS is minimized. The notion of requesting the content from the other users within a social group is attractive due to the fact of similar interests of all the users within a social group. This approach can be used to significantly improve the performance of edge computing-based AR for smart tourism.   
    \item We derived that novel and lightweight security mechanisms must be proposed for AR-based industrial applications. For example, an unauthorized user can access the remote live support system and interrupt the proper operation of the system, which can lead to serious consequences.
\end{itemize}

%%%%%%%%%%%%%%%%%%%%%%%%%%%%%%%%%%%%%%%%%%%%%
\begin{table*}[!t]
  \centering
  \caption {Classification, key contributions, and evaluation of the recent advances} \label{tab:recent_advances_evaluation1} 
  \begin{tabular}{|>{\centering\arraybackslash}m{2.5cm}|p{1.8cm}|p{7cm}|l|l|l|l|l|}
    \hline
   
    \multirow{2}{*}{Application} & \multirow{2}{*}{Reference}&  \multirow{2}{*}{Key contributions} & \multicolumn{5}{|l|}{Evaluation parameters} \\ 
    \cline{4-8} 
     &       &       & P1 & P2 & P3 & P4 & P5 \\
    \hline
    \multirow{3}{*}{Smart transportation}
 & Giang \textit{et al.}~\cite{smarttransportation2}  &\begin{itemize} \item Studied a fog assisted smart transportation applications requirements. \item Discussed state-of-the-art transportation applications. \item Identified research challenges in fog enabled smart transportation systems. \end{itemize} & \centering{\cmark} &\cmark & \xmark & \xmark  & \xmark \\ 
    \cline{2-8}
& Huang \textit{et al.}~\cite{huang2017exploring}  & \begin{itemize} \item Proposed a 5G SDN enabled vehicular network architecture. \item Discovered VNG using real data set. \item Proposed architecture considered sustainability, context-awareness, and security. \end{itemize}&\cmark & \cmark & \cmark & \xmark &  \xmark \\
  \cline{2-8}
& Premsankar \textit{et al.}~\cite{smartTransportation3}  &\begin{itemize}\item Proposed deployment of RSUs to jointly reduce deployment and transmission power cost. \item QoS is considered for vehicular applications during deployment of RSUs. \end{itemize} &\cmark & \xmark  & \cmark & \xmark & \xmark \\
     \hline 
    \multirow{4}{*}{Augmented reality}
 & Fernandez-Carames \textit{et al.}~\cite{augmentedreality3}  &\begin{itemize} \item Proposed a three layered edge computing and AR enabled architecture for industries. \item Presented a real time implementation of the proposed architecture.  \end{itemize} & \xmark  &\cmark  & \xmark & \cmark & \xmark    \\
  \cline{2-8}
& Taleb \textit{et al. }~\cite{taleb2017mobile}  &\begin{itemize} \item Proposed an architecture assisted by AR for high definition video streaming. \item Presented a shared memory concept for effective active service migration.
\item Validated the proposed architecture using a practical use case. \end{itemize} & \cmark &\xmark  &\xmark  & \cmark & \cmark  \\
  \cline{2-8}
& Schneider \textit{et al.}~\cite{augmented3}  &\begin{itemize} \item Proposed an architecture for remote live support. \item A video compression is adopted to reduce the bandwidth requirement. \end{itemize} & \xmark &  \xmark & \xmark & \xmark  &  \xmark \\
  \cline{2-8}
& Qiao \textit{et al.}~\cite{augmented4} &\begin{itemize}\item Proposed an edge assisted Web-based augmented reality. \item Discussed the deployment challenges of the proposed system.\end{itemize} & \xmark & \xmark & \xmark & \cmark & \xmark  \\
  \hline
  \multirow{3}{*}{Smart health-care}
 &  Muhammad \textit{et al.}~\cite{muhammad2018edge} &\begin{itemize} \item Proposed an edge computing assisted voice disorder assessment and treatment framework.\item The proposed framework provides interoperability. \end{itemize} & \cmark & \xmark & \xmark & \cmark & \cmark  \\
  \cline{2-8}
&  Rahmani \textit{et al.}~\cite{healthcare2}  &\begin{itemize} \item Proposed a three layer fog assisted framework for smart hospitals. \item Presented implementation of the IoT based remote health-care system as proof-of-concept. \end{itemize} &  \cmark& \cmark & \cmark & \cmark &  \cmark \\
  \cline{2-8}
&  Gia \textit{et al.}~\cite{healthcare3}  & \begin{itemize} \item Proposed a remote Health monitoring System for ECG along with joint analysis and notification. \item The fog assisted gate-ways also provides security by enabling restriction to un-authorized access. \end{itemize} & \cmark & \xmark & \cmark & \xmark & \cmark \\
 \hline
  \end{tabular}
\end{table*}
\subsection{Smart Health-care}
\setlength{\parindent}{0.7cm}Smart health-care is realized by having low cost and effective real-time health-care facilities ubiquitously.\par
\subsubsection{Smart Voice Disorder Treatment Framework}
\setlength{\parindent}{0.7cm}A voice disorder assessment and treatment framework enabled by edge computing and deep learning is proposed in \cite{muhammad2018edge}. Initially, smart sensors are used to collect samples of users voice, which are then processed on edge nodes and finally, sent to core cloud. Decision of automatic assessment is then sent by cloud manager to specialists for issuing a prescription to patients. A database used for training, testing and validation is the Saarbruken Voice Disorder (SVD) database. Prime limitation of their work lies in the usage of few disorders, such as vocal fold polyp, vocal fold paralysis, laryngitis, Kontakt pachydermia, hyper-functional dysphonia, functional dysphonia, and cordectomy, whose samples have sufficient amount in the database. \par
\subsubsection{Smart Hospitals}
\setlength{\parindent}{0.7cm}A three-layered fog based health-care architecture for homes/ smart hospitals was proposed in \cite{healthcare2}. The proposed architecture comprises of smart devices layer, edge layer, and cloud layer. The devices layer has medical sensors that captures the biomedical signals from the human body. These biomedical signals are further sent to the gateways through wireless technologies, such as ZigBee, Wi-Fi or Bluetooth. The cloud layer consists of cloud computing that performs data analytics. The edge layer consists of smart e-health gateways, that are distributed geographically and capable of supporting protocols for wireless communication and inter-device communication. In addition, the edge layer also performs local data processing, data filtering, data compression, data fusion, data analysis, local storage, and interoperability functions. \par
\subsubsection{Remote Electrocardiography Monitoring}
\setlength{\parindent}{0.7cm}In \cite{healthcare3}, a smart health-care monitoring system is proposed, that enable continuous remote monitoring of Electrocardiography (ECG) signals. The monitoring system consists of a cloud, fog assisted gateways, and sensor nodes. Purpose of sensor nodes is to acquire bio-signals (such as human temperature, respiration, and ECG) and contextual information (such as room temperature and humidity). Fog assisted gateway receives the signals from sensor nodes and processes it for further sending it to cloud. For example, pre-processing of the bio-signals are necessary for removal of noise and extraction of useful information from it. Furthermore, the fog assisted gateways have two storage databases, such as local and external user database. A local database is used to store the information that can be updated or edited by the administrators only. On the other hand, an external database is used for storage of bio-signals along with contextual information, such as room temperature and humidity. Data stored within the external database has synchronization with cloud servers. Additionally, the fog assisted gate-ways also provides security to restricts the unauthorized access.
\subsubsection{Lessons Learned: Summary and Insights}
We concluded that it is necessary to use federated learning for enabling intelligent edge-based smart health-care applications \cite{khan2019federated,pandey2019crowdsourcing}. Different health-care applications driven by machine learning are disease identification and diagnosis, drug discovery and manufacturing, better radiotherapy, outbreak prediction, crowdsourced data collection, and medical imaging diagnosis \cite{ML_health}. Traditional machine learning requires the migration of data from devices to a centralized location. However, this approach suffers from the serious privacy concerns. On the other hand, federated learning offers on-device distributed machine learning and thus, preserves user privacy which is of significant interest in smart health-care. Therefore, it is recommended to enable smart health-care applications via federated learning. However, the enabling of federated learning at the edge requires on-demand computational resources \cite{edgeAI}. Therefore, federated learning and edge computing must be used simultaneously to enable smart health-care applications.

\subsection{Smart Grids}
\setlength{\parindent}{0.7cm}A smart grid is a cutting edge technology to enable the transformation of the traditional grids to reduce utility costs and global warming using renewable energy resources, smart meters, and smart appliances.\par
\subsubsection{Smart Grid Data Management}
\setlength{\parindent}{0.7cm}Authors in \cite{kumar2016vehicular} proposed a four-layer architecture for data management based on Vehicular Delay-tolerant Network (VDTN) and edge computing for smart grids. First layer consists of mobile devices, such as Plug-in Hybrid Electric Vehicles (PHEV). The gateways and routers in the backbone network make second layer of the proposed architecture. Third layer has different types of servers, such as a certificate authority server, file server, and database server and fourth layer is made by a distributed cloud (infrastructure as a service). In a smart city environment, charging of PHEV batteries from the nearby charging station may result in many requests at charging stations. The proposed architecture is based on the use of VDTN for delivery of charging and discharging decisions. The authors compared the delay incurred using the traditional core-cloud infrastructure and edge computing enabled infrastructure and observed less incurred delay, more throughput, and less response time for edge computing.\par      
\subsubsection{Cloud–Fog–Based Smart Grid Model}
\setlength{\parindent}{0.7cm}In \cite{smartgrid3}, a three-layered framework named cloud-fog-based smart grid is proposed. The first layer is the end users layer, followed by the second fog layer, and the third cloud layer. The end layer comprises of buildings which have homes. Energy storage systems with renewable energy generation units are installed at every home. The information related to energy generation and consumption is sent to the fog layer. The fog layer that physically exists close to the consumers region, performs effective management of the network resource with low latency. Furthermore, the authors proposed hybrid artificial bee ant colony optimization (HABACO) algorithm for optimal allocation of virtual machines in the cloud-fog-based smart grid model. The authors considered two scenarios for performance evaluation of their cloud-fog-based smart grid model but did not consider the commercial scenario that has a significantly different nature than residential areas.\par
\setlength{\parindent}{0.7cm}A reference model for integration of fog computing and smart grids is introduced in \cite{smartgrid2}. The three-tier model consists of smart grid tier, fog tier, and cloud tier. The smart grid tier has the responsibility of enabling communication between different smart devices, such as smart meters, mobile devices, electrical vehicles, and smart appliances. The fog tier acts as an intermediate layer between the cloud and smart grid tier. Every fog server communicates with smart meters for the collection of data of their associated users. Additionally, the fog layer servers take the private data from the smart grid layer in encrypted form to ensure the privacy of the users. The fog server does not have the capability to decrypt the data and the key is only shared among the cloud and the users. The main purpose of the fog servers is to provide temporary storage and computation capability for the smart grids data. The prime limitation of their work lies in non-consideration of the challenge of interoperability, which is one of a most important issue among different power domain areas due to the usage of multiple semantic data models in a smart grid environment \cite{smartgrid_interoperability}.  \par
\subsubsection{Lessons Learned: Summary and Insights}
We concluded that privacy and security must be given primary importance in smart grids enabled by edge computing. Specifically, certain power plants such as nuclear and hydroelectric power plants must be provided with effective security and privacy mechanisms. A typical smart grid based on edge computing and cloud has multiple layers: Cloud layer, edge layer, and IoT layer \cite{gai2019permissioned}. Enabling security at the cloud is easier than the edge layer because of its centralized nature. On the other hand, edge computing-based power plants are based on distributed architectures, thereby more vulnerable to security attacks. For instance, consider energy trading in edge computing-based smart grid. To make edge computing-based smart grids more secure, we can use blockchain for energy trading at network edge between subscribers and energy suppliers. However, using blockchain in edge computing-based smart grids have significant challenges with respect to resource optimization (i.e., computational and communication resources) for reaching a consensus among blockchain nodes.

\begin{table*}
  \centering
  \caption {Classification, key contributions, and evaluation of the recent advances} \label{tab:recent_advances_evaluation2} 
  \begin{tabular}{|p{2.5cm}|p{1.8cm}|p{7cm}|l|l|l|l|l|}
    \hline
   
     \multirow{2}{*}{Application} & \multirow{2}{*}{Reference}&  \multirow{2}{*}{Key contributions} & \multicolumn{5}{|l|}{Evaluation parameters} \\ 
    \cline{4-8} 
     &       &       & P1 & P2 & P3 & P4 & P5 \\
    \hline
    \multirow{3}{*}{Smart grids}
 & Kumar\textit{et al.}~\cite{kumar2016vehicular} & \begin{itemize} \item Proposed a generalized architecture for data management based on VDTN using edge computing for smart grids. \item An energy efficient virtual machine migration utilizing load forecasting is also proposed. \end{itemize}& \xmark &\xmark &\cmark  & \xmark &\cmark \\
  \cline{2-8}
& Okay \textit{et al.}~\cite{smartgrid2}  & \begin{itemize} \item Proposed a model for integration of smart grid with fog computing. \item Evaluated the proposed model using qualitative use case.   \end{itemize}& \cmark & \cmark &\cmark & \xmark &\cmark  \\
  \cline{2-8}
&  Zahoor \textit{et al.}~\cite{smartgrid3} & \begin{itemize} \item Proposed a three layered cloud-fog-based smart grid framework. \item A hybrid artificial bee ant colony optimization algorithm is also proposed for virtual machines allocation. \end{itemize} & \cmark & \cmark & \cmark & \xmark  & \xmark \\
   \hline
    \multirow{2}{*}{Smart farming}
 & Ferrandez-Pastor \textit{et al.}~\cite{ferrandez2018precision} & \begin{itemize}\item Proposed a flexible layered IoT assisted PA architecture enabled by edge computing. \item Provides interoperability among different sub systems in the proposed PA architecture. \item Experimental validation of the proposed PA architecture is also provided. \end{itemize} & \cmark  &\cmark & \cmark &\xmark  &\cmark \\
  \cline{2-8}
& Caria \textit{et al.}~\cite{smart_agriculture2}  & \begin{itemize} \item Proposed a programmable Fog computing enabled architecture of smart farm for welfare of animals. \item Presented a Raspberry Pi assisted implementation of the proposed architecture. \end{itemize} & \xmark & \cmark &\cmark & \xmark  & \cmark\\
  \cline{2-8}
& Zamora-Izquierdo \textit{et al.}~\cite{smartfarming3}  & \begin{itemize} \item Proposed a three planar edge assisted PA architecture. \item NFV is exploited in edge layer to enable flexible generic control modules.\item The proposed architecture is validated using prototype implementation at real green-house in Spain.  \end{itemize}& \xmark & \xmark & \cmark & \xmark & \xmark \\
  \hline  
    \multirow{2}{*}{Smart buildings}
 & Vallati\textit{et al.}~\cite{vallati2016mobile}  & \begin{itemize} \item Proposed an edge computing enabled smart home architecture using LTE. \item Discussed the advantages and disadvantages pertaining to placement of edge severs at different positions, such as eNB, home, and UE. \end{itemize} & \cmark &\cmark &\xmark & \xmark &\cmark \\
  \cline{2-8}
&  Stojkoska \textit{et al.}~\cite{smarthome}  & \begin{itemize} \item Proposed a three tier IoT framework enabled by Fog computing for smart homes.   \end{itemize} & \xmark & \xmark& \cmark& \xmark &\xmark \\
   \cline{2-8}  
   \hline
\begin{comment}
     \multirow{1}{*}{Smart gaming}
 & Premsankar \textit{et al.}~\cite{smartgaming} &\begin{itemize} \item Discussed different edge computing architectures. \item Presented the use case of edge computing enabled mobile gaming. \end{itemize} & \xmark & \xmark& \xmark& \xmark &\xmark \\
 \hline
\end{comment}
   
\multirow{1}{*}{Smart management}
 &  Jia \textit{et al.}~\cite{smartManagement1} &\begin{itemize} \item Proposed an IMCS enabled by edge computing. \item Context-awareness, scalability, and sustainability are also exploited in the proposed system. \item Narrow-band IoT is used for communication due to its high energy efficiency, low cost, and support for massive number of manhole covers. \end{itemize} & \cmark & \cmark &\cmark & \xmark  & \xmark \\

 \hline
  \end{tabular}
\end{table*}

\subsection{Smart Farming}
\setlength{\parindent}{0.7cm}Smart farming is aimed to utilize advanced technologies for increasing productivity and reduce cost simultaneously \cite{smartfarming}. \par
\subsubsection{Smart Precision Agriculture}
\setlength{\parindent}{0.7cm}In \cite{ferrandez2018precision}, a distributed IoT-based smart framing infrastructure assisted by edge nodes is proposed to experience farmer with enhanced benefits. The proposed architecture consists of four levels, i.e., things layer, edge layer, fog layer, and communication layer. In things layer, a thing can be either actuators, controller or sensors, which are designed to meet production facility. The function of edge computing is to provide interoperability, storage, smart analysis, and prediction near the edge. Reason for using edge computing is due to the low latency requirements of control, analysis response, and monitoring sensors. Although authors considered interoperability but did not consider caching in the proposed framework. Caching can be utilized to further improve the performance of the smart framing \cite{cachingInPA}. In \cite{cachingInPA}, a joint latency and energy optimized communication were achieved through data caching algorithm in Precision Agriculture (PA) system. The data caching algorithm was based on the optimization of sleep-up/wake-up time of wireless sensor network nodes. Data requests of sensor nodes are analyzed in a cache prior to transmission. This prolongs the life of the sensor nodes and thus, improves the energy efficiency of the PA system.\par
\setlength{\parindent}{0.7cm}In \cite{smartfarming3}, the authors integrated PA with edge computing to enable real-time actions for better productivity. They proposed three planar architecture, such as the local plane, edge plane, and cloud plane. In a local plane, cyber-physical systems perform a collection of data through interaction with crop devices to enable real-time control actions. The second edge plane is responsible for performing tasks of the PA near the network edge to enhance reliability. The third cloud tier exists remotely to enable data analytics and smart management. The cyber-physical systems are connected to sensors and actuators, which are installed at the crop premises for the collection of temperature, humidity, electrical conductivity, and pH meter. All the cyber-physical systems have inter-connectivity with the Internet using multiple access networks technologies, such as digital subscriber line, optical fiber, and microwave radio links. In the edge plane, the tasks including energy and alarm management, greenhouse control, and nutrition control are performed. In the proposed architecture, the authors did not consider security which must be given proper consideration. Some malicious user might access the system and perform wrong activities, such as unnecessary alarms and windows opening.        \par
\subsubsection{Smart Animal Welfare Monitoring System}
\setlength{\parindent}{0.7cm}In \cite{smart_agriculture2}, a smart farm assisted by edge computing for the welfare of animals is proposed. The proposed architecture consists of an animal-centric system, environmental system, and farm controller. The farm controller can be implemented either in a centralized fashion or in a distributed fashion. A Raspberry Pi (R-Pi) which is a small sized micro-computer is used in the implementation of the environmental subsystem and animal-centric system. In system implementation, the environment R-Pi monitors the environment, whereas wearable R-Pi monitors the health of animals. A workstation that implements the farm controller performs the central tasks of the system. One of the most prominent advantages of the proposed architecture is its ability to send alarms to mobile. These alarms are use full to detect whether animals are ill or not, especially when the cow is pregnant which can be detected from its body temperature \cite{cowPregnancy}. The authors did not consider the robustness of the sensors network used for the collection of data which might suffer from the data loss due to battery depletion or other sources as observed in the \cite {wsnDataLoss}.\par
\subsubsection{Lessons Learned: Summary and Insights}
The lessons learned and derived future prospects from this section are as follows:
\begin{itemize}
    \item A typical precision agriculture system consists of several layers: Sensors layers, edge computing layer, and cloud layer. To enable efficient operation of smart agriculture, sensors layer require instant instructions and computational resource for processing of complex algorithms at the network edge. One way is to store frequently requested instructions at the remote cloud due its high storage capacity than edge computing servers. However, this will increase the delay and affects the system performance. Another way is to store these frequently requested instructions at the network edge. Therefore, we concluded that intelligent edge caching to store the frequently requested instructions by the sensors layer can be used in precision agriculture for performance enhancement. In a typical layered architecture of precision agriculture consisting of a sensor layer, edge server layer, and cloud layer, intelligent edge caching reduces the transmissions required by the sensors with a remote cloud for performing different operations. The instructions requests by the sensors layer are analyzed by the intelligent cache at the edge server before forwarding the request to the cloud. The requested instructions are provided instantly to the sensors if they are available in the intelligent edge cache otherwise the request is forwarded to the remote cloud. This type of transmissions reduction between the sensors and remote cloud due to intelligent caching significantly increases the overall throughput of the network and thus, improves the precision agriculture performance.    
    \item Similar to smart health-care, AR, and smart transportation, it is necessary to propose effective and lightweight security mechanisms for edge computing-based smart farming. A malicious user can access the system and perform some malicious activity, such as unnecessary alarms and windows opening in a precision agriculture scenario. 
\end{itemize}

\subsection{Smart Buildings}
\setlength{\parindent}{0.7cm}Smart buildings provide the users with advanced automated controls, i.e., smart meters, smart water control, smart plug-in hybrid electric vehicles (PHEV) charging, smart elevators, and smart security.\par
\subsubsection{IoT based Smart Homes}
\setlength{\parindent}{0.7cm}A smart home architecture is presented in \cite{vallati2016mobile}, that leverages edge computing and Long-Term Evolution (LTE) network. Different alternatives for placement of edge computing servers are considered that include:(a) placement of edge computing server at Evolved Node B (eNB), (b) placing the edge computing server at femtocells BS in a home, and (c) placing the edge computing server at User Equipment (UE). The positioning of edge computing server at eNB needs to cover a large number of far users and thus, require high computational power. On the other hand, placement of edge computing server near the UEs require less capacity due to its small coverage area. Simulation for two different scenarios, such as device-to-infrastructure (D2I) and D2D communication in IoT was performed using SimuLTE tool\cite{simlte}. \par
\subsubsection{Sustainable Smart Homes}
\setlength{\parindent}{0.7cm}A three-tier framework for smart homes based on IoT and edge computing is proposed in \cite{smarthome}. Three tiers of the framework are a smart homes tier, a nanogrid tier, and a microgrid tier. The smart home tier consists of household items with interfaces to enable wireless communication. All of data at smart homes is collected at the sink, which has ability of local processing and storage. In the nanogrid tier, sinks of different smart homes communicate with each other. To enable communication between sinks, different typologies, such as mesh, cluster or star can be used. In the microgrid tier, sinks of different smart homes communicate with utility through gateways. IoT devices have limited energy, therefore edge computing enabled gateways can be used to provide the computation services to smart homes. Major limitation of the framework lies in lack of considering security which poses significant challenges to smart homes operation. A malicious might access the IoT-based smart home objects and operates it in an irregular way \cite{yaqoob2019internet}.\par
\subsubsection{Lessons Learned: Summary and Insights}
We derived the following lessons from this section:
\begin{itemize}
    \item The key requirement of interoperability (discussed later in section~\ref{sec:req}) is derived from this subsection. A large number of heterogeneous IoT devices must perform seamless interaction with each other to enable edge computing and IoT-based smart homes.
    \item We derived that it is necessary to propose novel forensic tools and models for edge computing-based smart environments. A typical smart home consists of a wide variety of IoT devices that interact with each other and edge computing server. These IoT devices and edge servers are highly susceptible to security attacks. Therefore, it is necessary to develop novel forensic models and tools for edge computing-based smart environments to enable investigation of the attacks. 
\end{itemize}

\subsection{Smart Management}
\setlength{\parindent}{0.7cm}In the smart city era, information and communication technologies are incorporated to enable smarter urban infrastructure management for improving overall user QoE. In \cite{smartManagement1}, an Intelligent Manhole Cover Management System (IMCS) that leverages edge computing is proposed for avoiding accidents due to the manhole cover. The proposed IMCS comprise of three parts: perceived recognition technology, network transmission technology, and information processing technology. In perceived recognition technology, a Radio-Frequency Identification (RFID) tags and sensors are attached with manhole to facilitate active communication. In the second part, narrow band-IoT and the Internet are used for communication between the manholes and end users. The third part consists of an edge computing enabled intelligent management system that allows immediate decision making. The authors considered context-awareness, scalability, and sustainability in the design of IMCS, but did not consider security.      
\subsubsection{Lessons Learned: Summary and Insights}
We concluded that the edge computing can serve as a PaaS for running of compute-intensive algorithms required for smart management. Apart from that, sensors used in smart management generates an enormous amount of data which motivates the use of effective machine learning techniques for making the management system smarter. On the other hand, different cities have different context-awareness features, such as location of the sensors and their surrounding environment. Therefore, it is necessary to consider context-aware machine learning models for smart management systems in cities rather than considering a general machine learning model. Based on context-awareness, machine learning for smart cities can be either global context-aware or local-context aware. The former is trained for a set of users located in a large geographic area, whereas the latter covers the users of similar interests reside at the network edge.    

%\subsection{Lessons Learned: Summary and Insights}
%\setlength{\parindent}{0.7cm}From this section, we learned that edge computing is a viable for realizing real-time smart cities advancements. We derived five parameters, such as context-awareness, scalability, sustainability, caching, and security, from recent advances literature for rigorous evaluation. We categorized the recent advances into different application areas and summarized their key contributions in table~\ref{tab:recent_advances_evaluation1}-\ref{tab:recent_advances_evaluation2}. We then derive that security has considerable importance among the other factors and thus must be implemented in the design of smart applications. For instance, consider autonomous driving and smart health-care enabled by edge computing; malicious user can access the system and information, which might result in accidents. Other than security, the incorporation of context-awareness in edge computing can improve the performance of emergency response systems.          
\section{Taxonomy}
\label{sec:taxonomy}
\begin{figure*}[ht]
	\centering
	\captionsetup{justification=centering}
	\includegraphics[width=18cm, height=15cm]{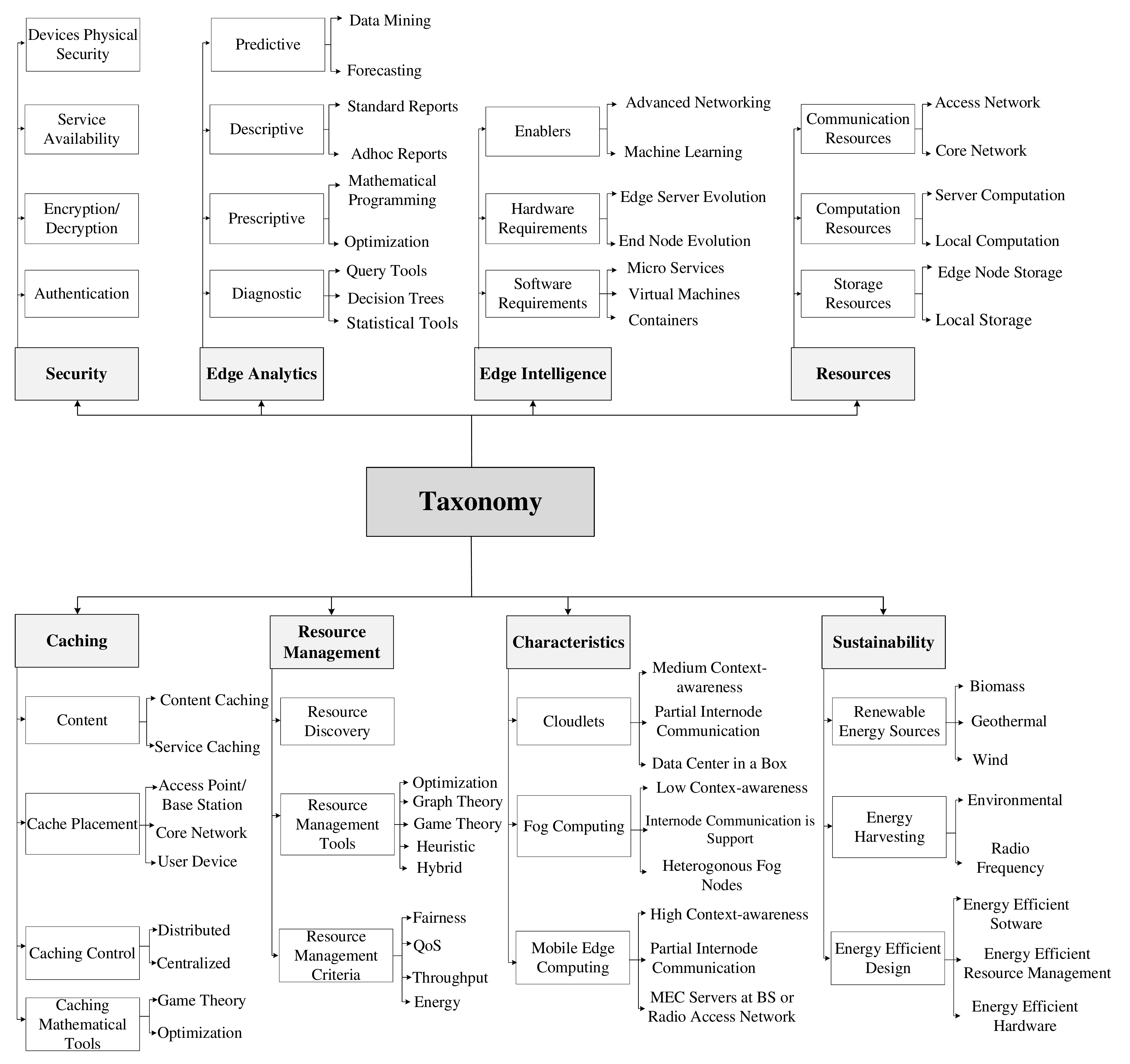}
	\caption{Taxonomy of edge computing enabled smart cities.}
	\label{fig:taxonomy}
\end{figure*}

\setlength{\parindent}{0.7cm}Figure~\ref{fig:taxonomy} elucidates the taxonomy which is devised using edge intelligence, edge analytics, resources, caching, characteristics, sustainability, security, and resource management. Further details are provided in the following subsections.
\subsection{Characteristics}
\label{sec:characteristics}
\setlength{\parindent}{0.7cm}Edge computing paradigms generally appear in three different formats, such as cloudlets, fog computing, and mobile edge computing \cite{MECSurvey3}. All of these formats use the same notion of migrating the computing resources from the cloud to the edge of the networks. However, there exist unique differences among them in terms of context-awareness, inter-node communication, and the position of the computing resources. Mobile edge computing has the highest context-awareness due to their availability of user device location, network load, and network capacity at the edge servers \cite{contextawrenesscomp}. Cloudlets have the lowest context-awareness due to their standalone architecture, which is connected to a cloud. Fog computing has context-awareness between that of mobile edge computing and cloudlets. The ability of inter-node communication somewhat improves the context-awareness level of fog nodes. Other than context-awareness, inter-node communication has more support in fog computing than both cloudlets and mobile edge computing. \par
The key feature of robust edge computing architecture for smart cities is derived from this section for future research. A typical edge computing-based smart environment consists of a wide variety of players that interact with each other. The commonly used players are edge servers, IoT devices, remote cloud, and communication infrastructure. Every technology has distinct specifications and requirements. Therefore, it is necessary to propose novel robust architectures that enable the integration of edge computing with other emerging technologies to offer different smart services. One way to achieve robust operation is through fail-over and redundancy techniques. A fault affected edge server can be replaced by another backup server to continue the operation. However, this approach increases the cost of the system. On the other hand, different critical smart applications (i.e., autonomous driving, delivery UAV control, and smart industrial manufacturing) require ultra-reliability. Therefore, there must be a trade-off between the robustness and cost of the system. 

\subsection{Edge Analytics}
\label{sec:analytics}
\setlength{\parindent}{0.7cm}Edge analytics deals with the analysis of data near the network edge, and then sending back these results to the edge devices. This has a significant impact on reducing the delay incurred when sending data to a centralized cloud for analysis. In smart cities many devices, including smart sensors, and other smart devices from a variety of real-time smart applications, including augmented reality, smart health-care, smart industrial control, smart transportation, and smart surveillance, generate a huge amount of data. To enable such real-time applications, it is necessary to perform data analytics at the edge rather than at the centralized remote cloud. In the future, fog nodes, cloudlets, and mobile edge computing servers will be designed with the ability to perform big data analytics. \par
\setlength{\parindent}{0.7cm}Similar to cloud computing, edge analytics can be descriptive, diagnostic, predictive, and prescriptive. Descriptive analytics is a simple class of analytics that enables users to transform big data into small chunks with useful information. It is preferable to use descriptive analytics when understanding is required at an aggregate level. This allows users to learn from past events, which can be from either not long (e.g., 2 min) or from a long time ago (e.g., 10 weeks). Examples of descriptive analytics are percentage changes, average, sum, among others. Predictive analytics allow the use of machine learning and statistical modeling for analyzing historical data and predicting future trends. Sentimental analysis is an example of predictive analysis. Prescriptive analytics allows performing near optimal or optimal actions in accordance with the predicted or actual data. Prescriptive analytics include mathematical programming (such as integer programming, non-linear programming, and linear programming.) and optimization. On the other hand, diagnostic analytics is based on examining to understand "Why somethings happened?". For diagnostic analytics at the edge, query tools, decision trees, and statistical tools can be used.  
\subsection{Resources}
\label{sec:resources}
\setlength{\parindent}{0.7cm}Edge computing converges information technology and telecommunication services to enable computation of resource-intensive tasks at the network edge. To carry out the above process, we need three types of resources, such as computation resources, communication resources, and storage resource \cite{MECSurvey2}. Computation resource can be either a local device or server. A user offloads either all components of an application (binary offloading) \cite{binartoffloading1,binartoffloading2,binartoffloading3} or some components of an application to the edge server (partial offloading) \cite{partialoffloading1,partialoffloading2}. A complication might arise in partial offloading when the task components are dependent on each other, and making it difficult to separate the task components for local execution and execution at server. Apart from computation resource there are two types of communication resources, that are required to enable edge computing in smart cities. One is the access network, which allows connectivity of the end-user devices to the edge servers, and the other is a core network communication resource, which allows transmission of data from the edge server to cloud server at the time of insufficient computation resources at the edge server. Different from computation and communication resources, storage resources can be either at a local device or at an edge.       
\subsection{Resource Management Strategies}
\label{sec:RMS}
\setlength{\parindent}{0.7cm}In smart cities enabled by edge computing, the key resources are the massive number of smart devices, edge servers, and communication infrastructure. To enable edge computing-based smart environments, it is necessary to effectively manage the available resources. The main aspects of resource management include resource discovery, resource management tools, and resource management criteria. Fairness \cite {fairness1,fairness2}, QoS \cite{Qos1,Qos2}, throughput \cite{throughput1}, energy \cite{energy1,energy3,energy4,energy5}, among others, can be utilized as criteria to design resource management schemes. Other than resource management criteria, we can use optimization-based schemes \cite{optimization1}, game theory-based schemes \cite{gametheory2}, heuristic \cite{huristic1}, graph theory-based schemes \cite{MEC4, wang2019coupling}, and hybrid schemes (which jointly utilize optimization and game theory) \cite{hybrid1}, for resource allocation. An optimization-based technique can be either dynamic programming, linear programming, convex optimization, or integer programming. On the other hand, game theory-based schemes can be based on either cooperative games or non-cooperative games. \par
Herein, we derived the use of computation replication in edge computing enabled smart environments. Typically, a task or sub-task in edge computing is migrated toward a single edge computing node. The edge computing node sends back the computed result to the requesting user. On the other hand, a wireless channel is specified by random variations that significantly degrades the signal quality. Therefore, it is necessary to use computation replication in edge computing enabled smart services. In computation replication, multiple copies of the same task or a sub-task are transmitted to different servers. After the computation of the task by all the edge servers, the result is transmitted back to the user using a diversity scheme to further decrease the download time and improve the signal quality. Although the computation replication increases robustness and decreases download time, it suffers from increased upload time due to generally low up-link bandwidth nature. Therefore, a trade-off must be made between the robustness and up-link upload time. Additionally, we derived that there must be effective mobility management schemes for edge computing enabled smart environments. A set of IoT nodes involved in different smart environments (i.e., smart transportation) can have significant mobility. Therefore, it is necessary to propose novel mobility management schemes for edge computing-based smart environments to enable seamless connectivity with edge computing servers.

\subsection{Edge Intelligence}
\label{sec:intelligence}
\setlength{\parindent}{0.7cm}Edge intelligence will transform edge computing analytics capabilities to the next level by enabling self-learning solutions. The three fundamental aspects of edge intelligence are key enablers, hardware requirements, and software requirements \cite{edge_intelligence_1}. Edge intelligence can be enabled through advanced network capabilities and machine learning. Other than key enablers, the edge server must be evolved to yield a virtualized and truly dynamic micro data center. Additionally, end devices must be evolved to handle intelligent operations. Apart from hardware evolution, the key goal of edge computing is to isolate the functionality from the hardware that is provided by services form. Such isolation has the advantage of being deployable anywhere in edge computing enabled smart cities. The concept of micro-services can be used for isolation, whose deployment requires the same run-time environment at all deployment locations. Both virtual machines and containers can be used to run  micro-services. However, containers offer less overhead than virtual machines. Moreover, containerization is also preferable in the cloud for deployment of micro-services. Therefore, at the edge it is also preferable to use containerization to provide a uniform run-time environment. We derived the open research challenge of \emph{Intelligent Edge} from this section that is further discussed along with possible guidelines in section.~\ref{intelligentedge}. Edge analytics can play an important role in future smart cities. Massive numbers of smart devices and sensors produce a huge volume of data, out of which only a small portion of data is useful. This requires data filtering and intelligent instant processing, which can be enabled by intelligent edge-based AI.   

\subsection{Caching}
\label{sec:Caching}
\setlength{\parindent}{0.7cm}Caching is temporarily storing the network contents to avoid repeated transmissions. In a typical communication network, the back haul-links suffer from congestion due to repeated requests of the same contents. For instance, in the Internet traffic, the video stream is the major source of traffic and as counts for 54\% of all traffic, which is estimated to reach 71\% in 2019 \cite{cisco2016global}. Therefore, it is necessary to store the video contents temporarily at different locations to avoid repeated transmissions. Edge computing with caching can be used to enable resource-intensive applications, such as smart industries, smart transportation, augmented reality, smart tourism, to name a few \cite{cachinginSmartIndustries,cachinginSmartTransportation,cachinginTransportation,cachinginAR}. Caching in edge computing enabled smart cities has four main aspects such as cache location, cache contents, caching control, and caching mathematical tools. Caches can also be placed either at the core network or at the Access Point (AP) depending the on the requirements. Cache placed at the core network requires a high capacity due to the fact that multiple access points will be connected to it. Apart from that, a cache can be placed at the AP or at the UE. Different from cache placement, caching control may be either centralized \cite{cachecentralized} or distributed \cite{cachedentralized}. Furthermore, control structure and mathematical tools for caching can be based on machine learning, optimization, and game theory \cite{gametheoryincaching,cachinggame}. The challenge of \emph{Intelligent caching} which is discussed in detail along with possible guidelines in section.~\ref{intelligentcaching}, is derived from edge caching for different smart applications. Intelligent caching based on AI and machine learning is expected to offer a considerable decrease in latency for real-time applications. Moreover, it increases the core network throughput and offers a solution to mobility-aware caching.

\subsection{Sustainability}
\label{sec:Sustainability}
\setlength{\parindent}{0.7cm}Sustainability refers to the use of energy efficient designs and renewable energy resources to reduce the overall carbon footprint. In future smart cities, densification of devices and servers is expected, which in turn will result in energy limitations. Therefore, sustainable developments are required. Sustainability in edge computing can be achieved by using renewable energy sources \cite{renewableenergy1}, energy harvesting \cite{harvesting1,harvesting2,harvesting3}, and energy-efficient design \cite{energyefficientdesign1}. Renewable energy sources include biomass, geothermal, and wind energy. Energy efficient design has three important aspects, such as energy efficient software, energy-efficient hardware, and energy efficient resource management. Other than energy efficient design and renewable energy sources, energy harvesting deals with the extraction of energy from external sources for further use in the operation of small devices, such as sensors and wearable devices. Energy can also be harvested either from the environmental sources, such as wind, solar, and thermal, or from radio frequency sources. There are significant variations in the harvested energy from both natural sources and radio frequency sources \cite{harvesting1}. Therefore, it is imperative to design hybrid systems that utilize both harvested energy and energy from the grid. A hybrid system initially must entirely use harvested energy, and then later use energy from the grid in instances when the harvested energy level becomes lower than the required energy for operation. The use of hybrid energy sources to enable sustainable operation is derived in this section. Hybrid energy sources make use of both harvested energy and grid energy on demand. First, such a system utilizes the available harvested energy. The grid energy is used only when the harvested energy level falls below the required energy level of the application. \par 
\begin{figure*}[!ht]
	\centering
	\captionsetup{justification=centering}
	\includegraphics[width=16cm, height=15cm]{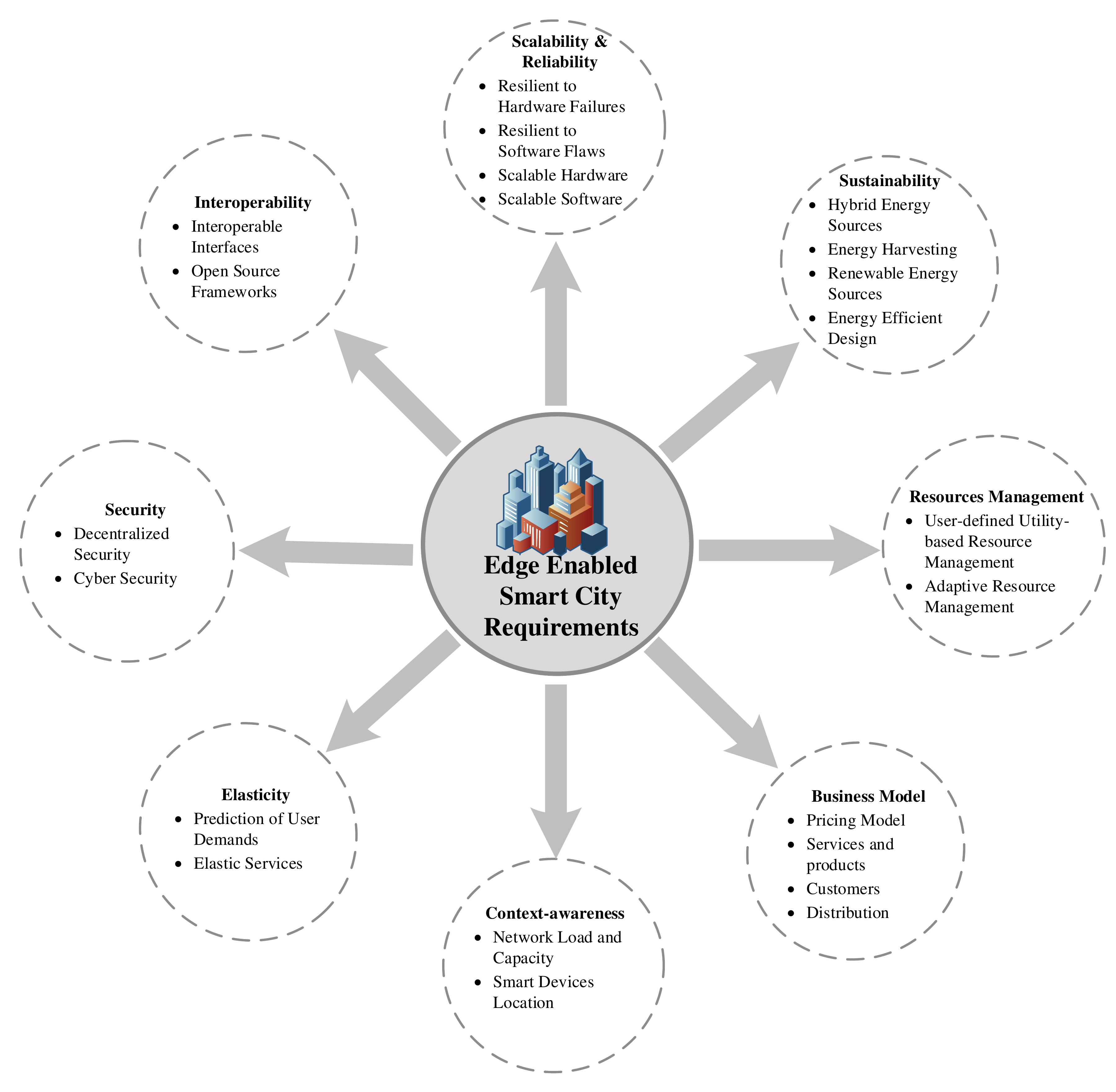}
	\caption{Edge computing enabled smart cities requirements}
	\label{fig:requirements}
\end{figure*}
\subsection{Security}
\label{sec:security}
\setlength{\parindent}{0.7cm}Edge computing infrastructure exhibits novel security risks and challenges due to its distributed nature in contrast to a centralized cloud \cite{edgesecurity1,edgesecurity2,MECSurvey3,edgesecurity4}. In edge computing, numerous players, such as virtualization platforms, distributed and peer-to-peer systems, and wireless networking technologies are involved \cite{edgesecurity5}. To enable security, we must protect all of these players from attacks by implementing diverse security mechanisms, whose orchestration is often very complex \cite{MECSurvey3}. In addition, the security schemes must be autonomous due to the decentralized nature of edge computing. The prominent benefit of autonomous and distributed security mechanisms is their lower latency compared to centralized security mechanisms.\par
Generally, edge computing security risks can be classified into denial of service attacks, man in the middle situations, physical damage, and privacy leakage \cite{MECSurvey3}. An attacker can easily access end user devices or connect to the edge servers, and then install malicious software on it. Therefore, it is very difficult to secure end-user devices and edge servers physically due to their geographically distributed nature. On the other hand, in man in the middle attacks, a malicious user takes control of the wireless network, followed by traffic injections, whereas privacy leakage refers to unauthorized access of a users information stored at the edge data center. Denial of service attacks refer to wireless jamming. Both denial of service and privacy leakage have a limited scope than a man in the middle attacks. Therefore, to avoid man in the middle attacks, we must use effective and lightweight encryption/decryption schemes \cite{encryption_decryption1, encryption_decryption2,encryption_decryption3}. Furthermore, effective authentication protocols must be devised for edge enabled smart cities to ensure user privacy \cite{authentication1,authentication2,authentication3,authentication4, authentication5, authentication6}.  In \cite{authentication4}, a novel authentication scheme based on certificateless public key cryptosystem has been proposed to provide secure transport of messages between mobile devices acting as edge computing nodes and autonomous connected vehicles. Another paper studied radio frequency identification (RFID) based authentication scheme for edge computing \cite{authentication5}. The authors proved that authentication schemes based on states-limited synchronization are not able to well defend the denial of service attacks. Moreover, a privacy-preserving authentication protocol with low-complexity is proposed. On the other hand, a clustering-based physical-layer authentication scheme has been proposed in \cite{authentication6}. The proposed scheme combined clustering and lightweight symmetric cipher with channel state information to provide mutual authentication between edge nodes and IoT devices. Moreover, the proposed scheme is validated via a real-time multiple-input multiple output-orthogonal frequency division multiplexing (MIMO-OFDM) system. The authors proved that their proposed scheme can resist again the spoofing attacks, replay attacks, and integer attacks. 
\section{Requirements}
\label{sec:req}
\setlength{\parindent}{0.7cm}This section outlines and explains core requirements to turn the vision of edge computing enabled smart cities into reality. These requirements (illustrated in figure~\ref{fig:requirements}) include scalability and reliability, resource management, interoperability, sustainability, elasticity, context-awareness, security, and business models. A more detailed description of the requirements is given below. \par
\subsection{Scalability and Reliability}
\label{Scalability&reliability}
\setlength{\parindent}{0.7cm}How will edge computing enabled smart cities infrastructure simultaneously respond to a gigantic increases in the numbers of end-user devices and system failures?\par  
\setlength{\parindent}{0.7cm}In smart cities, we will witness massive increases in the numbers of actuators, smart sensors, and other smart devices, subsequently resulting in unprecedented growth in data traffic. To enable the resource-intensive and latency sensitive smart city applications, edge computing provides a solution. To create a scalable edge computing infrastructure, we must consider reliable performance for an increasing number of end users, dynamic smart devices, and diverse networks \cite{bringingedge}. Scalability in edge computing infrastructure can be achieved by the addition of new service providing points. Addition of new service points require scalable hardware and software. On the other hand, prominent challenges of reliability in edge computing infrastructure might arise because of its dynamic nature, software flaws, hardware flaws, and network failure. To cope with the system failures, we must use edge computing based infrastructures within smart cities which are resilient to hardware and software flaws. Other way is to use duplicate service providing points. However, this will cause a significant increase in the cost of a system. Moreover, this approach adds complexity to the system because of its distributed nature. As the edge computing is mainly intended to enable strict latency applications. Therefore, the addition of duplicate nodes must be immediate, which is very challenging due to their distributed nature. Other ways to ensure reliability include the use of robust hardware, fault-tolerant networking protocols, and robust software. \par
The open research challenge of collaborative edge computing is derived from this section. The authors in \cite{verticalhorizontalcolloboration} and \cite{smartcollaboratingnetworking} considered collaborative edge computing to offer scalability and reliability. Edge computing servers are characterized by a low computing capacity that must be utilized efficiently. Additionally, there are significant variations in smart device traffic. To cope with variations in the massive amount of smart devices traffic along with scalability and reliability constraints, the use of collaborative edge computing is a viable solution.\par
\subsection{Context-awareness}
\label{sec:awareness}
\setlength{\parindent}{0.7cm}How does one enable smart cities with mission-critical applications such as smart health-care services and smart transportation services using edge computing?\par
To answer this question, the edge server must be context-aware (such as it must know the locations of smart devices) to enable computation with strict latency \cite{contextawareness1,contextawareness2,contextawareness3,contextawareness4}. Moreover, the edge server requires knowledge of network load and capacity to enable resource-efficient communication with smart devices. For example, if an accident occurs in a smart transportation system, then it is necessary to respond as soon as possible for first aid. The arrival of first aid services is heavily dependent on the user emergency call. A severely injured user might not be able to call for emergency service. On the other hand, context-aware edge servers at RSU can enable reporting of emergencies in a timely manner with very low latency. Therefore, it is evident that context-awareness is one of the key requirement in the design of edge computing enabled smart cities infrastructure.\par
\subsection{Resource Management}
\label{Resource Management}
\setlength{\parindent}{0.7cm}How does one effectively manage computation, communication, and storage resources between different players in smart cities to optimize the overall system performance? \par
\setlength{\parindent}{0.7cm}To answer this question, we must devise resource management schemes that consider energy \cite{energyresource1,energyresource2,energyresource3, energyresource4}, latency \cite{latency1}, traffic handling capacity \cite{MEC4}, and user-defined utility \cite{utility1} to optimize the overall performance of the system \cite{resourcemanagement1}. However, it is very difficult to simultaneously consider all of these factors when designing a resource management scheme. There are three main categories of resources in edge computing enabled smart cities: computing resources, communication resources, and storage resources \cite{resourcemanagement2}. To enable cost-efficient operation with better QoS, resource management algorithms must be adaptive and perform management according to an applications nature. For example, more computing resources must be given to a self-driving car than AR to enable faster execution of the self-driving car functions to avoid accidents \cite{carlatency1}. On the other hand, the throughput requirement of AR is larger than for self-driving cars \cite{ARbandwidth}. Therefore, a trade-off must be made between computation and radio resource allocation.  \par
We derived that there must be novel resource optimization schemes based on AI for edge computing-based smart environments. It is expected that a typical $5$G node will have more than $2000$ configurable parameters \cite{edgeAI}. It is recommended in \cite{8382166} to use AI for data link and physical layer. In addition, deep reinforcement learning and federated learning based framework has been proposed in \cite{edgeAI} for managing communication, caching, and computation at the network edge. Therefore, it is recommended to use federated learning at the network edge for resource management in different smart environments. \par
\subsection{Sustainability}
\label{Sustainability}
\setlength{\parindent}{0.7cm}How does one develop edge computing enabled smart cities without losing QoS and adding pollution to the environment?\par 
\setlength{\parindent}{0.7cm}Sustainability is another important requirement when designing edge computing infrastructures for smart cities. In smart cities, the densification of end-user devices and edge computing servers is expected. Consequently, this puts significant energy limitations on the overall smart city infrastructure design. Therefore, it is imperative to tackle the challenge of high energy consumption without degrading QoS. The three main aspects are related to sustainability of the edge computing enabled smart cities: energy efficient design \cite{energyefficientcaching1, energyefficient1,energyefficient2, energyefficient3}, use of renewable energy sources \cite{RenweableEnergy1,sustainabilityEDGE,RenweableEnergy3}, and energy harvesting \cite{enviornmentalEH1, enviornmentalEH2, RFEH1, RFEH2 }. The edge computing enabled smart cities infrastructure is designed to reduce energy consumption by employing a novel architecture design and communication technologies. Numerous ways to improve the energy efficiency of edge computing system design include energy-efficient caching \cite{energyefficientcaching1}, energy-constrained offloading and sleeping \cite{energyefficient1}, and energy-aware computation offloading and user association \cite{energyefficient3}. Furthermore, energy-efficient wireless networking technologies, such as  IPv6 over low-power WPAN (6LoWPAN) and ZigBee can be used to enable sustainable operation of a system \cite{energyefficientcommunication}. In \cite{8752525}, Gai \textit{et al.} considered cognitive wireless communication that uses reinforcement learning and edge computing to enable energy efficient operation. Other than energy efficient design, an edge computing ecosystem that utilizes energy from the fossil fuels-based power plants will add to the carbon footprint. Therefore, an edge computing infrastructure must utilize energy from renewable energy sources such as wind, solar, and hydropower. Apart from that, energy can be harvested either from environmental sources, such as wind and sun \cite{enviornmentalEH1, enviornmentalEH2} or radio frequency sources, such as from radio frequency transmitters and interference signals \cite{RFEH1,RFEH2,InterferenceEH}. Energy harvesting offers significant advantages but suffers from random variations in both environmental and radio frequency energy. Therefore, we must use hybrid sources that also utilize harvesting energy and grid energy on demand if the harvested energy level falls below the required energy. From this section, we derived the adoption of high-performance memory requirement for sustainable and reliable operation. The edge servers are intended for use in multiple diverse applications to serve massive number of users. Moreover, edge servers are also expected to enable support for computationally expensive machine learning algorithms. Therefore, use of high-performance memory is one of the promising solution for enabling sustainable edge computing enabled smart environments.\par
\subsection{Elasticity}
\label{Elasticity}
\setlength{\parindent}{0.7cm}How does one enable smart citizens with services in an elastic manner to avoid both under and over-utilization of resources?\par
\setlength{\parindent}{0.7cm}Unprecedented increases in the numbers of sensors and other smart devices are expected in smart cities. This will impose limitations on the available resources, such as computing, communication, and storage resources. Therefore, it is necessary to design systems that permit efficient utilization of the resources. Elasticity in edge computing enables to avoid over-allocation and under-allocation of computing resources. This can be realized in smart city infrastructure by providing elastic services based on the allocation of resources with the ability to grow and shrink dynamically according to demand \cite{elasticity1,elasticity2,elasticity3}. For elasticity, it is necessary to properly predict user demand via edge service providers prior to assignment of computational resources, such as bandwidth, processing power, and storage. Other than prediction, edge computing-based smart services must be elastic. In \cite{elasticity4}, a proactive scheme is presented to enable auto-scaling by predicting future resources based on past data. Auto-scaling techniques can be either vertical, horizontal, or hybrid \cite{elasticity5}. Vertical scaling refers to re-sizing of resource allocation on the same node, whereas horizontal scaling is re-sizing of resource allocation among a cluster of nodes cluster. A hybrid approach combines both vertical and horizontal schemes.  \par
The open research challenge of network slicing-based smart city is derived from this section. In a typical smart city, we will expect a wide variety of players: Edge service providers, cloud service providers, telecommunication infrastructure providers, among others, that interact with each other to offer different smart services. Enabling successful interaction between these service providers and smart citizens that jointly optimize users QoS and service providers profit can be made possible via network slicing. The network slicing providers will buy resources from the physical infrastructure providers and sell it back to smart citizens to increase the providers profit and improve users experience.\par
\subsection{Interoperability}
\label{Interoperability}
\setlength{\parindent}{0.7cm}How does one operate the massive number of heterogeneous devices in edge computing enabled smart cities?\par 
\setlength{\parindent}{0.7cm}Interoperability is the ability of different systems to understand and utilize each others functions. Similar to IoT, interoperability is one of the most important requirements in the design of the edge computing enabled smart city infrastructure. The challenge of interoperability in edge computing enabled smart city infrastructure arises due to massive number of heterogeneous devices running different protocols. To enable seamless operation, edge computing architectures must be able to provide support for interoperability. Interoperability can be implemented in smart cities by exploiting inter-operable interfaces and open source frameworks.\par
\subsection{Privacy and Security}
\label{privacy&security}
\setlength{\parindent}{0.7cm}How does one preserve user privacy and data from attacks by malicious users?\par 
\setlength{\parindent}{0.7cm}Security in edge computing infrastructure has two main aspects: devices cyber security and physical security. Cyber security refers to protection of the network, computing infrastructure, and data from attacks. Specifically, data security is the protection of data from unauthorized access, whereas privacy can be considered as legislation (a set of rules) that defines the levels of protection. Privacy places restrictions on the authorization of users regarding data access and the purpose of data usage. Apart from privacy, security must be properly addressed in smart cities. At the core of edge computing paradigms, there are many different technology enablers, i.e., virtualization, wireless networks, peer-to-peer, and distributed systems. Therefore, enabling physical security of edge devices is necessary and very challenging (due to the distributed nature of the edge nodes). An unauthorized user can easily physically access any of the edge devices and install malicious software. Other than physical security, effective cryptographic techniques can be used to provide improved security against the attacker. Furthermore, security mechanisms in edge computing ecosystems can be either centralized or decentralized based on continuous dependence on the centralized infrastructure. Centralized security requires continuous availability of the centralized infrastructure. Although this has the advantage of easier management it is more vulnerable to failure by an attack on the centralized server. On the other hand, a decentralized security mechanism has the advantages of reduced delay and being less vulnerable to failure by attacks. However, such a mechanism has a higher complexity than centralized algorithms due to its distributed nature.  \par
We derived from this section that the use of blockchain in edge computing-based smart cities is necessary. A blockchain is a type of distributed ledger technology (DLT) that offers several striking features of decentralization, transparency, democracy, enhanced security, and immutability \cite{SC2}. These promising features of blockchain made its use favorable in secure and transparent sharing of data by citizens for enabling different smart applications. However, blockchain suffered from the scalability barrier with an increase in the number of transactions \cite{8624417}. To cope with this challenge of scalability, edge computing is a promising solution by providing on-demand computational resources and storage. However, the integration of edge computing and blockchain in a smart city further brings challenges that must be resolved.\par
\subsection{Business Model}
\label{sec:buisnessmodel}
\setlength{\parindent}{0.7cm}How does one increase revenue of the edge computing service providers while maintaining user QoS?\par 
\setlength{\parindent}{0.7cm}Smart cities enable a plethora of horizontal and vertical use cases that have diverse requirements. Additionally, the cities involve different players, such as end users, cloud providers, edge computing service providers, and communication infrastructure providers, interplay with each other. The already available business models for cloud computing are not feasible because of the distributed nature of edge computing. Therefore, to commercially deploy the edge computing infrastructure, the existing service providers will have to adopt new pricing models to generate more revenue while maintaining a better QoS. We derived from this section that novel business models must be developed for edge computing enabled smart cities. The already developed business models for cloud computing are insufficient for edge computing enabled smart cities due to their distributed architecture. \par
\section{Case Studies}
\setlength{\parindent}{0.7cm}This section presents recently reported synergies and case studies of edge enabled smart cities. A summary of these case studies listing the goals, the organization involved, deployment status, and country is given in the section. \par
\begin{comment}

\begin{table*}[]
\caption {Summary of the case studies} \label{tab:casestudies} 
\begin{center}
\begin{tabular}{lp{5cm}p{1.5cm}p{4.5cm}l}
\toprule\toprule
\textbf{Case studies  } & \textbf{Key objectives} & \textbf{Deployment status}& \textbf{Organizations involved} & \textbf{Country} \\ \midrule
Barcelona smart city & \begin{itemize}\item Traffic flow management \item On-demand Connectivity \item Security and real-time analytics \item Efficient power monitoring \item Event based video \item Access control and cabinet telemetry  \end{itemize} & Deployed  & \begin{itemize} \item Barcelona City Council \item Prismtech \item i2cat \item Technical University of Catalonia \item Supercomputing Center \item Schneider Electric \item Plat.one \iteme Sensefields \item CISCO \end{itemize}  & Spain \\ \midrule
\#SmartME               &  \begin{itemize}  \item Smart parking \item Smart airport \item Smart taxi \item Smart lighting control \item Smart energy management \item Smart tour management \item Smart pothole detection \end{itemize}    & Deployed  & \begin{itemize} \item University of Messina \end{itemize} &   Italy  \\\midrule
WATCH                  &   \begin{itemize} \item Smart surveillance  \end{itemize}   &  2020  &  \begin{itemize} \item Future Intelligence Ltd, London \item London South Bank University  \end{itemize}      & United Kingdom \\ 
\bottomrule \bottomrule              
\end{tabular}

\end{center}
\end{table*}

\end{comment}
\subsection{Barcelona smart city}
\setlength{\parindent}{0.7cm}Barcelona smart city is a joint project of Barcelona City Council, Prismtech, i2cat, Technical University of Catalonia, Barcelona Supercomputing Center,  Schneider Electric, Plat.one, Sensefields, and CISCO to realize smart services \cite{barcelonasmartcity,barcelona1}. The project partners focused on five use cases that include power monitoring/element control, access control and cabinet telemetry, event-based video, traffic management, and on-demand connectivity. To realize a smart city, installation of large number of IoT driven devices is required, further creating operational and space problems. In this regard, the number of deployed cabinets is more than 3000 for Barcelona smart city realization. On the other hand, fog computing is validated to enable real-time decision making, power supplies for autonomous operation, guaranteed access control, local data analysis, and execution of complex algorithms for the assistance of vehicles flow.\par
\subsection{\#SmartME}
\setlength{\parindent}{0.7cm}\#SmartME is a project provisioned to transform the city of Messina into a smart city \cite{smartme1}. The project was initiated by a group of researchers from the University of Messina. The key goals of the \#SmartME project are establishment of a smart city infrastructure that allows all citizens to contribute to infrastructure through hardware sharing. The \#SmartME framework consists of three-layer: an application layer, Stack4Things layer (for fog computing platform implementation), and city layer. The Stack4Things layer is a framework developed at the University of Messina with the objective of enabling the administrators to manage IoT devices without taking care of their physical location. Main features of the Stack4Things include object virtualization, overlay network of things, remote control and customization, and fog orchestration. The applications developed on the top of the \#SmartME infrastructure include \#SmartME Parking, \#SmartME lighting, \#SmartME pothole, \#SmartME  airport, \#SmartME taxi, \#SmartME art, and \#SmartME trashcan.\par
\subsection{WATCH}
\setlength{\parindent}{0.7cm}The \textit{Wide Smart Safe, Robust and Resilient Smart Cities Application Using Fog Computing} (WATCH) project is funded by INNOVATE UK \cite{watch1}. It was started on Nov 01, 2017 and is expected to be completed by 2020. The lead participants of this project are Future Intelligence Ltd. London, and London South Bank University, United Kingdom. The goal of this project is to exploit novel micro-data center and telecommunication nodes to provide joint storage resources, local processing, and networking to enable support for user applications in smart cities. Further, they will utilize software-defined networking and Network Function Virtualization (NFV) to enable the formation of interconnected devices islands. This resulted in fogs, which are actually small scale clouds at the network edge (i.e. edge computing). The main objective of this project is improvement of smart surveillance by leveraging edge computing.\par
\subsection{Lessons Learned: Summary and Insights}
\setlength{\parindent}{0.7cm}This section presented the recently reported case studies of edge computing enabled smart cities. We discussed the objectives, organizations involved, and their current deployment status. From this section, we concluded that the development of high-performance simulation tools for implementation and validation of edge computing based smart cities applications is required to be done. For instance, the IoTIFY is an online cloud based network simulator that enables simulation of smart waste management, smart parking spots, smart street lights, smart traffic signals, and smart transportation \cite{iotify}. We must develop novel high-performance simulators to test edge computing and other emerging technologies in smart city environments. \par      
\section{Open Research Challenges}
\label{sec:researchchallenges}
\setlength{\parindent}{0.7cm}This section presents open research challenges in the realization of edge computing in smart cities. Furthermore, a summary of these challenges, their causes, and possible guidelines is given in table~\ref{tab:challenges}. 
\subsection{Intelligent Caching}
\label{intelligentcaching}
\setlength{\parindent}{0.7cm}In smart cities, connectivity of billions of smart IoT devices is expected. According to CISCO, 507.9 ZB data will be generated by smart IoT devices in 2019 \cite{hassan2018role}. This rapid rise in data has resulted in a major bottleneck that requires higher back-haul data rates. To tackle the dilemma of back-haul congestion, caching can be used which deals with the storage of popular contents at different locations in a network to avoid repeated transmissions \cite{cachingsurvey}. Additionally, caching enable different real-time IoT-based smart city applications by reducing latency and energy consumption \cite{cachinginsmartcities}. On the other hand, caching has two main aspects: network type and caching strategy. Network type deals with cache positioning, such as cache enabled macro cellular networks \cite {cachecentralized}, cache enabled heterogeneous networks \cite {cacheenabledhetnet}, cache enabled D2D networks \cite {cacheenableD2D}, and cache enabled cloud radio access networks \cite{cacheneabledCRAN}. The caching strategy has two main phases, such as cache decision phase and content delivery phase. \par
\setlength{\parindent}{0.7cm}The main challenges that exist in intelligent caching at the edge include the cold-start user problem \cite{cachingcontent2} and security and privacy \cite {security3}. Intelligent caching based on machine learning uses information and data from all users within its range. However, the mobility of the users poses many challenges and it is very difficult to satisfy this requirement for intelligent caching. To cope with user mobility, mobility-aware hierarchical caching was proposed in \cite {cachinginSmartTransportation}. Furthermore, we can leverage reinforcement learning for effective scheduling in mobility-aware caching \cite{Reinforcementbasedcaching1}. Additionally, intelligent cache decisions are based on the future content popularity, whose computation is considerably challenging. The prediction of content popularity using deep learning has been considered to enable intelligent cache decision making \cite{cachingcontent1,cachingcontent2,cachingcontent3,cachingcontent4}.               
\begin{table*}
% \makegapedcells
\caption {Summary of the research challenges and their guidelines} \label{tab:challenges} 
  \centering
  \resizebox{\textwidth}{!}{
  \begin{tabular}{p{3cm}p{6.5cm}p{6.5cm}}
    \toprule 
   
    \textbf{Challenges} &  \textbf{Causes} & \textbf{Guidelines} \\
   \midrule
      \textbf{Intelligent caching} & \begin{itemize} \item Data sparsity due to mobility  \item  High back-haul congestion  \item  High latency in content delivery   \end{itemize} & \begin{itemize}  \item Deep learning assisted cache content prediction \item Reinforcement learning based mobility-aware caching \end{itemize}  \\
   \midrule
    \textbf{Collaborative edge computing} & \begin{itemize} \item Edge server inherent low computation and storage capacity constraints \item Profit and QoS dependency on size, number, and coverage area of edge servers \item Absence of high scalability and reliability  \end{itemize} & \begin{itemize} \item Horizontal edge computing collaboration \item Vertical edge computing collaboration \item Inter-domain collaboration \item Intra-domain collaboration \item Smart collaborative networking  \item Social trust based incentive design for collaboration \end{itemize} \\
 \midrule
     \textbf{Cooperative and sustainable load balancing}  &  \begin{itemize} \item Under-loaded scenarios  \item Over-loaded scenarios \item Substantial energy wastage \item Profit margin reduction  \end{itemize}  & \begin{itemize} \item Novel game theory assisted incentive mechanisms  \item Adaptive cooperation mechanisms \item Novel authentications based load balancing \end{itemize} \\
\midrule
    
   \textbf{Intelligent edge } & \begin{itemize} \item Absence of instant big data analytics  \item High latency in instant and autonomous decisions  \item  Semantic non-interoperability  \end{itemize}  &   \begin{itemize}  \item AI based predictive and diagnosis analytics  \item Machine learning based semantic inter-operable interfaces  \item Novel privacy regulation rules for personal data sets      \end{itemize}\\
\midrule
      
\textbf{Network slicing} &  \begin{itemize} \item Diverse requirements of smart city applications    \item Lack of effective resource management of multiple service providers \item Slice security challenges   \end{itemize} & \begin{itemize} \item Separation of control plane from data plane \item End-to-end slice management and orchestration  \item SDN-based orchestrator security  \item  Adaptive service function chaining  \end{itemize} \\
 
\midrule
 
  \textbf{Cyber and physical security} & \begin{itemize} \item Privacy leakage \item Denial of service  \item Service manipulation \item Multiple trust domains  \end{itemize} & \begin{itemize} \item Lightweight authentication schemes  \item Blockchain-based authentication schemes \item Certificateless public key cryptosystem-based authentication scheme \item SDN and NFV assisted security  \item Attribute based access control \item Federated capability-based access control  \end{itemize}  \\

\bottomrule 
\end{tabular}
}
\end{table*}
\subsection{Collaborative Edge Computing}
\setlength{\parindent}{0.7cm}Virtually enabled edge computing servers are characterized by low computation and storage capacity constraints \cite{MECservercapacityconstraints1,MECservercapacityconstraints2}. They use containers and virtual machines to enable different smart applications. Apart from that, edge computing servers can leverage dimensional parameters, such as server number, server capacity, and server operation area to optimize the overall performance in terms of QoS and operator profit. On the other hand, a tremendous increase in the number of smart devices is expected in the future. Consequently, this poses significant challenges pertaining to the scalability and QoS of smart city infrastructure. To enable scalable and resource-optimized operation of edge computing in 5G enabled smart cities, we can leverage collaborative edge computing. Collaborative edge computing aims to form a collaboration space between edge computing servers in close vicinity. Moreover, this enables the maximization of resource usage, latency minimization, and reduction of back-haul traffic. Collaboration in edge computing enabled smart city infrastructure can be categorized into horizontal and vertical collaboration \cite{verticalhorizontalcolloboration}. Horizontal collaboration involves cooperation among different edge computing nodes that exist simultaneously in the same edge computing layer. Vertical collaboration involves collaboration among different layers, such as the smart devices layer, edge computing layer, and remote cloud computing layer. Although vertical collaboration may add additional latency to the tasks that are offloaded to the edge layer, it offers significant advantages. One of these advantages is the capability to process flooded deep learning queries for enabling different smart applications by providing enhanced computing power. The edge computing layer generally cannot handle flooded deep learning queries due to its low-computational power \cite{han2019convergence}. Therefore, the flooded deep learning queries must be processed at the remote cloud. Another advantage of the vertical collaboration between the edge layer and cloud layer is the preprocessing of raw data at the edge layer before sending it to the cloud layer for further analysis.\par
\setlength{\parindent}{0.7cm}The major challenges that exist in collaborative edge computing include collaboration space formation, social trust-based incentive policies, cooperation policies, inter-domain cooperation, intra-domain cooperation, smart collaborative networking, and mobility management. Collaboration space formation must be adaptive and based on resource availability. Furthermore, within the collaboration space, collaboration schemes must be designed that minimize back-haul bandwidth usage subject to the constraints of server resource availability, tasks computation deadlines, and local computation capabilities. Apart from formation of the collaboration space, we must design social trust-based incentive policies and cooperation policies. Within a collaboration space, only edge computing nodes with social trust can share resources with each other based on incentives in the form of payment. A coalition game was used in \cite{trustbased}  to design pricing incentive based on trust for collaboration in ultra-dense networks. Additionally, mobility must be managed properly within a collaboration space to enable efficient resource usage. The design of collaboration algorithms regarding the mobility of resource deficient nodes, such as autonomous vehicles and UAVs, is challenging. \par
On the other hand, we must design novel algorithms for intra-domain and inter-domain collaboration among edge computing nodes. Inter-node collaboration occurs between mobile edge computing servers, cloudlets, and fog nodes, whereas, intra-domain collaboration involves similar edge computing nodes, such as mobile edge computing nodes, fog nodes, or cloudlets. Enabling intra-domain collaboration is easier than intra-node collaboration due to the homogeneous nature of edge computing nodes. Apart from inter-domain and intra-domain collaboration, smart networking technologies must be designed to enable latency sensitive and energy optimized collaboration. In \cite{smartcollaboratingnetworking}, the design of a smart internet architecture allowing smart collaboration was proposed. Furthermore, deep learning and game theoretic approaches can be used to enable smart devices, edge servers, and other networking devices to perform traffic flow control collaboratively \cite{Deeplearningbasedcontroltraffic}.      
\subsection {Cooperative and Sustainable Load Balancing}
\setlength{\parindent}{0.7cm}Smart device data exhibit significant temporal and spatial traffic variations resulting in possible congestion at edge computing enabled AP/BS \cite{loadbalancing1}. Consequently, this causes performance degradation pertaining to higher latency. Moreover, offering edge computing servers with dynamic loads causes under-loaded and overloaded scenarios. As a result, substantial QoS performance degradation and profit loss of edge service providers will occur. Additionally, the smart cities infrastructure requirement of high scalability is also adversely affected by under-utilization and over-utilization. Hence, it is necessary to perform adequate load balancing. On the other hand, edge computing servers consist of hardware resources, such as main memory and secondary storage which are virtualized \cite{loadbalancing2}. Virtual machines are then assigned to requested applications. Every virtual machine operation is characterized by two states, the idle state and the active state. The energy consumption of the idle state is approximately 60\% that of the active state. Therefore, it is necessary to avoid under-utilization of edge computing servers to enable sustainable operation.   \par
\setlength{\parindent}{0.7cm}Numerous challenges exist for cooperative and sustainable load balancing, including novel pricing incentive design, joint minimization of task dropout and task transfer, and secure authentication. In cooperative load balancing, the load balancing between different edge computing servers involves negotiation between the involved players. To enable cooperation between different edge computing servers, we need effective incentive mechanisms. Design of the incentive mechanism depends on the nature of the edge server strategy, such as an open strategy or sealed strategy. In an open strategy, the edge servers are allowed to share their resource information, whereas the resource information of the edge servers is only available to service providers in a sealed strategy. Game theory can be utilized in the design of an incentive mechanism to enable cooperative load balancing in edge computing \cite{icentivegame1}. The actions of the three main players of the edge computing ( i.e., service providers, users, and edge servers) can be used to model various game theory parameters, such as the action set and preference set. Apart from incentive mechanism design, the cooperation mechanism must be adaptive and offer the ability to trade-off between task dropout and task transfer \cite{cooperativeMechanism1}. Furthermore, security must be considered in the design of cooperative and sustainable load balancing \cite{cooperativesecurity1}.      \subsection{Intelligent Edge Computing}
\label{intelligentedge}
\setlength{\parindent}{0.7cm}A massive amount of data will be generated by smart devices and sensors for real-time smart city applications that require instant processing. Most of these data contain noise and only a small portion of it is useful. The challenge is how to separate large useful data sets (such as big data) from noisy data. A viable solution to this challenge is the use of AI in the edge. Intelligent edge computing can perform data filtering to yield only useful data before sending to a remote cloud. Moreover, it can also offer real-time processing of data. Intelligent edge computing offers real-time analytics that enable us to answer questions, such as \emph{Why are some things happening} and \emph{What should be done}. Additionally, it enables process analysis and prediction of future trends. More specifically, the use of instant analytics at the edge is necessary to enable big data analytics. On the other hand, semantic interoperability between different edge computing enabled systems is one of the major bottlenecks regarding the  exchange of information to enable effective training of the machine learning algorithms \cite{semanticinteroperability1}. Intelligent edge computing is a way of providing interoperability among non-interoperable systems. Additionally, intelligent edge computing offers security enhancement by enabling smart applications to avoid sending data to remote clouds for predictive analytics. \par
\setlength{\parindent}{0.7cm}The challenges relevant to developing intelligent edge computing based on AI include its high computational power, complex data acquisition and storage, user privacy, and interoperability. The execution of traditional AI algorithms requires high computational power. However, the computational power at the edge is lower than that of the remote cloud. Therefore, it is necessary to devise low-complexity learning algorithms. Apart from its high computational power, data acquisition and storage of smart devices and sensors in various smart environments is also challenging. The humongous data sets used for AI algorithm validation require high storage capacity. In addition, the data set might contain noisy or missing data, which must be handled properly. Another challenge in training AI algorithms is the integration of large data sets. Integrating and synthesizing large data sets from the different players in a smart environment for effective AI algorithm training has a very high computational complexity. To cope with this issue, an inter-operable system is a solution that enables systems to interact with each other and share data. In \cite{semanticinteroperability1}, an edge computing-based semantic gateway has been proposed to enable inter-operability in health-care systems. To enable semantic interoperability machine learning-based interoperable interfaces can be used. On the other hand, maintaining user privacy while utilizing data sets for training AI algorithm pose challenges. For instance, consider an edge computing enabled smart health-care system with patient personal data. The privacy of users must be maintained via effective security mechanisms. Moreover, effective privacy regulation must be devised to ensure the user privacy.  
\subsection{Network Slicing}
\setlength{\parindent}{0.7cm}In smart cities, a wide range of players including cloud server providers, edge computing service providers, telecommunication services providers, and IoT service providers, will interplay with one another. The goals of these service providers are to increase profit, whereas, users want to improve their QoS. To enable easier and efficient resource management among users and service providers, the concept of network slicing was introduced \cite {slicing1,slicing2, slicing3, slicing4, tun2019wireless}. Network slicing exploits software-defined networking (SDN) and NFV to separate the control plane from the data plane \cite{slicing2}. The concept of network slicing is based on the creation of logical networks on top of a physical network infrastructure.\par
\setlength{\parindent}{0.7cm}To enable network slicing in smart cities, it is imperative to tackle the challenges, such as end-to-end slice management and orchestration, slices security, and adaptive service function chaining. A network slicing architecture consists of different layers: a service layer, resource layer, and network slice instance layer \cite{networkslcingarchitecture1}. End-to-end slice management and orchestration spans all three layers and handles the creation and management of slices. Moreover, it deals with translation of all the smart city use cases into slices with its associated network functions. To enable effective network slicing, game theoretic, deep reinforcement learning, and auction theory-based schemes can be used \cite{gametheoryslicing1,gametheoryslicing2, actionbasedslicing1}. Apart from management and orchestration of slices, slice security is another key challenge in realization of network slicing in smart cities. The presence of numerous authorities and multiple stakeholders in network slicing pose several novel security challenges. Network slicing security involves challenges of multiple operator resource sharing security and SDN-based orchestrator security. The slices defined for different use cases have unique security requirements. For instance, a smart health-care slice has stronger security concerns than the infotainment slice. Therefore, we must consider different security mechanisms for individual slices. Overall, security must be ensured during resource sharing of the infrastructure. However, network wide-security has different nature than of individual slices security. Therefore, the development of novel and effective security policies for network resource sharing and individual slices is necessary.\par
\setlength{\parindent}{0.7cm}On the other hand, SDN-based orchestrator attacks involve unauthorized access by a malicious user. An authorized user can access only a single network orchestrator but still alter the network resources of multiple operators. Therefore, meticulous attention must be given to SDN-based orchestrator security. Apart from SDN-based orchestrator security, another core challenge for network slicing is adaptive service function chaining. The key driver of network slicing, NFV allows implementation of virtual network functions on virtual machines. To deliver a service, a service function chain can be considered that leverages one or more virtual network functions in a sequence. The design of an adaptive service function chaining mechanism that jointly reduces the network deployment cost and maximizes user QoS is necessary. In \cite{SFCmatchinggame1}, traffic-aware and energy-aware virtual network function placements are considered. A solution based on a matching game has been proposed to enable efficient placement of virtual network functions.  
\subsection{Cyber and Physical Security}
\setlength{\parindent}{0.7cm}In edge computing enabled smart cities, the connectivity of a massive number of smart devices is expected, which can be prone to serious security attacks. The different areas prone to security attacks include smart devices, network infrastructure, edge computing servers, and core infrastructure. These security attacks can be either physical or cyber attacks. Physical security in edge enabled smart cities is difficult to maintain due to the distributed nature of edge computing. A malicious user can easily access the edge nodes and perform malicious activity. Therefore, enabling edge computing with high cybersecurity is of primary importance.\par
\setlength{\parindent}{0.7cm}The challenges that exist in enabling effective security of edge computing enabled smart cities includes network security, effective authentication, and access control. Authentication offers the ability to determine user validity by sharing identification information between two parties. Traditional cryptography-based authentication schemes can be used to devise authentication schemes for edge computing \cite{security1}. However, the high computational complexity associated with these authentication techniques limits its application to edge computing. Indeed the computational resources are more limited in edge computing than cloud computing. Therefore, we must devise novel light weight authentication schemes with low complexity \cite{martinez2019enhanced, manzoor2019multi}. In \cite{security2},  a Lightweight Anonymous Mutual Authentication scheme for ntimes Computing Offloading (LAMANCO) was proposed for anonymous authentication between IoT devices and edge devices. A certificateless public key cryptosystem-based authentication scheme for edge computing enabled autonomous driving has been proposed in \cite{authentication4}. Other than that, we can use energy-aware RIFD-based authentication schemes for edge computing-based smart environments \cite{authentication5}. In \cite{wazid2019design}, a novel secure key management and user authentication scheme called as SAKA-FC has been presented. The key advantage of using SAKA-FC is the use of lightweight operations, such as bitwise exclusive-OR and a one-way cryptographic hash function. In \cite{kaur2019blockchain}, a blockchain-based authentication scheme has been proposed to offer the key features of trust, anonymity, and decentralization. Other than authentication, network security is another important aspect of smart city infrastructure. Emerging communication technologies such as Sigfox, LoRaWAN, ZigBee, 6LowPAN, WAVE, and 5G used in smart cities have their own security protocols. However, there is room for further research regarding network security in the context of edge computing. Session keys are negotiated using the distribution of credentials and novel and effective schemes are needed to distribute these credentials. Furthermore, SDN andNFV can be used to enable secure networking. Specifically, they can be employed for isolation of different types of traffic \cite {MECSurvey3}. They allow for traffic diversion from insecure devices towards secure devices.  \par
\setlength{\parindent}{0.7cm}Similarly, access control in edge computing must be given considerable attention. Any edge devices after authentication, with certain privileges, can access the virtualized edge servers and misuse them. Additionally, the existence of multiple trust domains associated with multiple service providers in a single ecosystem pose challenges for developing effective access control schemes. Attribute-based access schemes can be used to restrict access control in edge computing \cite{security3, security4}. However, the major downside of attribute-based schemes is the high complexity associated with the decryption phase. In \cite{security5}, an efficient access control scheme (namely CP-ABE) was proposed for fog computing with outsourcing capability. In \cite{8350294}, Gai \textit{et al.} proposed a Dynamic Privacy Protection (DPP) model based on dynamic programming that is used to enable privacy for mobile devices. The DPP model offers significant maximization of privacy weights and thus, might be suitable for use in several edge computing enabled smart environments in the future. Additionally, a file hierarchy attribute-based encryption scheme was proposed in \cite{security6} to enable fine-grained access control for fog computing. On the other hand, \cite{security7} proposed federated capability-based access control for IoT to offer effective access control with high scalability compared to attribute-based schemes.\par
\section{Conclusions and Future Prospects}
\label{sec:conclusions}
\subsection{Conclusions}
\setlength{\parindent}{0.7cm}Edge computing is a promising computing paradigm to enrich smart cities with instantaneous computing and storage resources. Typically, cloud computing is employed to offer smart devices with computation and storage resources. However, inherent delay of cloud computing has paved the way for migrating the computing and storage resources from a centralized remote location to the edge of the network. On the other hand, real-time smart city applications require instant analytic services. To enable these real-time applications, the use of edge computing is required. However, implementation of edge computing in smart cities poses significant challenges.\par
\setlength{\parindent}{0.7cm}In this survey, the adoption of edge computing in smart cities is comprehensively studied. To this end, we first presented the evolution of edge computing, discussing the gradual evolution of computing technologies towards edge computing. In addition, the technologies involved and the benefits of different computing paradigms are also presented. Second, significant recent advances are presented and a rigorous evaluation is performed using different assessment parameters. Apart from recent advances, we have classified the literature and devised a taxonomy according to different parameters, such as edge analytics, edge intelligence, resources, caching, resource management, characteristics, sustainability, and security. Additionally, numerous requirements for enabling smart cities via edge computing are proposed. We also reported a few case studies of enabling edge computing in smart cities. Furthermore, numerous open research challenges are discussed in detail, and causes and possible guidelines to cope with these challenges are summarized. \par  
\begin{table}[!t]
\caption {List of abbreviations} 
\label{tab:abbreviations} 
  \centering
  \begin{tabular}{ll}
    \toprule
   \textbf{Abbreviation} & \textbf{Word} \\\midrule
   AP & Access Point \\ \midrule
   AR & Augmented Reality \\ \midrule
   ARPA & Advanced Research Projects Agency Network \\ \midrule
   AI & Artificial Intelligence \\ \midrule
   BS & Base Station\\ \midrule
   CAGR & Compound Annual Growth Rate\\ \midrule
   CIoT   & Consumer Internet of Things\\ \midrule
   DARPA & Defense Advanced Research Projects Agency \\ \midrule 
   D2D  & Device-to-Device\\ \midrule
   ECG & Electrocardiography\\ \midrule
   eNB & Evolved Node B\\ \midrule
   ETSI & European Telecommunications Standards Institute\\ \midrule
   %ICT & Information and Communication Technology\\ \midrule
   IaaS & Infrastructure-as-a-Service\\ \midrule
   IoT & Internet of Things\\ \midrule
   IIoT & Industrial Internet of Things \\ \midrule
   IAR & Industrial Augmented Reality\\ \midrule
   IMCS & Intelligent Manhole Cover Management System\\ \midrule
   IoS & iPhone Operating System \\ \midrule
   LTE & Long Term Evolution \\ \midrule
   MCC & Mobile cloud Computing\\ \midrule
   NBIoT & Narrow-band Internet of Things\\ \midrule
   NFV & Network Function Virtualization\\ \midrule
   PHEV & Plug-in Hybrid Electric Vehicles\\ \midrule
   PaaS & Platform-as-a-Service \\ \midrule
   PA & Precision Agriculture\\ \midrule
   QoE & Quality of Experience \\ \midrule
   QoS & Quality of Service\\ \midrule
   R-Pi & Raspberry Pi\\ \midrule
   RSU & Road Side Unit\\ \midrule
   SaaS & Software-as-a-Service \\ \midrule
   SDN & Software Defined Networking\\ \midrule
   RFID & Radio-Frequency Identification\\ \midrule
   UE & User Equipment\\ \midrule
   UAV & Unmanned Aerial Vehicle\\ \midrule
   VDTN & Vehicular Delay-tolerant Network\\ \midrule
   VNG & Vehicular Neighbor Group\\ \midrule
   VR & Virtual Reality\\
  \bottomrule
  \end{tabular}
\end{table}
\subsection{Future Prospects}
\setlength{\parindent}{0.7cm}We expect network slicing to be widely adopted in developing smart cities in the foreseeable future. Currently, millions of devices are deployed in cities. In the future, the addition of trillions of smart devices and sensors is expected \cite{7955906}. All of these devices generate data traffic which requires instant analytic services. Edge computing can enable instant analytics by providing on-demanding computing capabilities with extremely low latency. Furthermore, novel communication technologies with high throughput are required for data transmission between end devices and edge servers. On the other hand, the development of smart cities must be sustainable and reliable. To enable smart cities through novel technologies, different players such as telecommunication network operators, edge computing service providers, cloud service providers, and the IoT service providers will interplay with each other. One main issue is how to effectively enable such interplay to maximize profit and improve QoS? Network slicing is an answer, which can be leveraged to improve QoS and increase profit. This allows creating multiple logical networks (called slices) on top of a physical infrastructure of different service providers \cite{slicing2, NScity2, NScity3, NScity4}. The key enablers of network slicing enabled smart cities are SDN and NFV. SDN offers separation of the control plane from the data plane, and NFV allows the use of generic hardware for the implementation of different network functions. Separating the control plane from the data place offers easier network management and more flexibility of adding new functionalities to the network. \par
\setlength{\parindent}{0.7cm}Network slicing-based smart cities will offer a wide variety of slices for different smart services. Different references, such as \cite{NSsmartcityapp1, NSsmartcityapp2, NSsmartcityapp3, NSsmartcityapp4} have considered network slicing-based smart city advancements. These slices have diverse requirements and must share the same physical infrastructure resulting in a cost-efficient operation. Allocation of an end-to-end network for a typical slice would be expensive. Therefore, it is necessary for multiple slices to coexist simultaneously. However, resource management of resources between co-existing slices is challenging, thus requiring novel resource management schemes. Different organizations are also attempting to enable smart applications via network slicing. For instance, consider reference \cite{NSsmartcityapp1}, which introduces a network slicing-based testbed using edge computing at Hamburg port. Additionally, many organizations are working on the AutoAir project in the UK to produce a testbed based on edge computing and network slicing for testing connected and autonomous vehicles. To sum up, edge computing and network slicing will be the dominant solution for enabling future smart cities.\par
%\section*{Acknowledgements}
%This work was partially supported by an Institute for Information communications Technology Promotion (IITP) grant funded by the Korean government (MSIT) (No. 2015-0-00557, Resilient/Fault Tolerant Autonomic Networking Based on Physicality, Relationship and Service Semantic of IoT Devices) and the MSIT (Ministry of Science and ICT), Korea, under the ITRC (Information Technology Research Center) support program (IITP-2018-2013-1-00717) supervised by the IITP (Institute for Information & communications Technology Promotion).
\section*{Appendix}
\subsection*{Abbreviations}
See table~\ref{tab:abbreviations}.

\bibliographystyle{IEEEtran}
\bibliography{Database}
\begin{IEEEbiography}[{\includegraphics[width=1in,height=1.25in,clip,keepaspectratio]{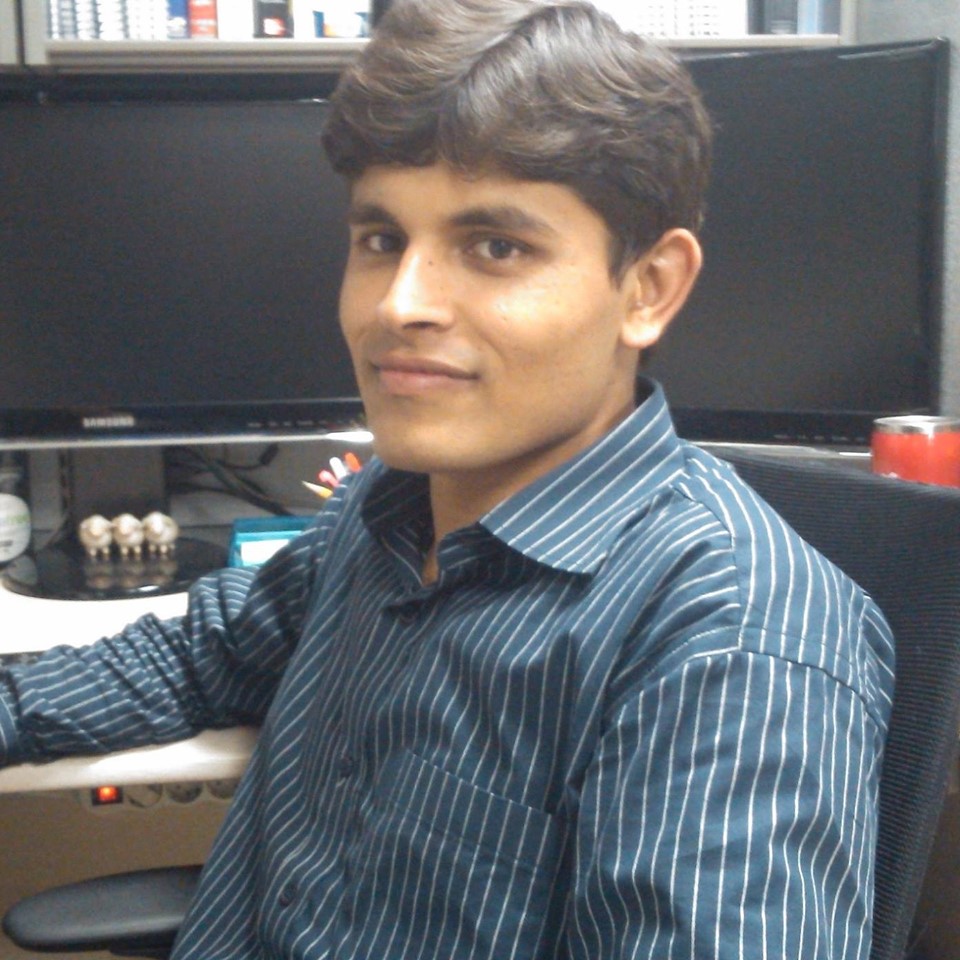}}]{Latif U. Khan} is currently pursuing his Ph.D. degree in Computer Engineering at Kyung Hee University (KHU), South Korea. He is working as a leading researcher in the intelligent Networking Laboratory under a project jointly funded by the prestigious Brain Korea 21st Century Plus and Ministry of Science and ICT, South Korea. He received his MS (Electrical Engineering) degree with distinction from University of Engineering and Technology (UET), Peshawar, Pakistan in 2017. Prior to joining the KHU, he has served as a faculty member and research associate in the UET, Peshawar, Pakistan. He has published his works in highly reputable conferences and journals. His research interests include analytical techniques of optimization and game theory to edge computing and end-to-end network slicing.  \end{IEEEbiography}
\begin{IEEEbiography}[{\includegraphics[width=1in,height=1.25in,clip,keepaspectratio]{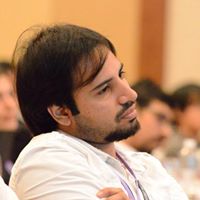}}]{Ibrar Yaqoob} (S'16-M'18-SM'19) is a research professor with the Department of Computer Science and Engineering, Kyung Hee University, South Korea, where he completed his postdoctoral fellowship under the prestigious grant of Brain Korea 21st Century Plus. Prior to that, he received his Ph.D. (Computer Science) from the University of Malaya, Malaysia, in 2017. He worked as a researcher and developer at the Centre for Mobile Cloud Computing Research (C4MCCR), University of Malaya. His numerous research articles are very famous and among the most downloaded in top journals. He has reviewed over 200 times for the top ISI- Indexed journals and conferences. He has been listed among top researchers by Thomson Reuters (Web of Science) based on the number of citations earned in last three years in six categories of Computer Science. He is currently serving/served as a guest/associate editor in various Journals. He has been involved in a number of conferences and workshops in various capacities. His research interests include big data, edge computing, mobile cloud computing, the Internet of Things, and computer networks. 
\end{IEEEbiography}
\begin{IEEEbiography}[{\includegraphics[width=1in,height=1.25in,clip,keepaspectratio]{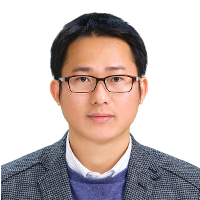}}]{Nguyen H. Tran}(S'10-M'11-SM'18) is currently working as a senior lecturer in the School of Computer Science, The University of Sydney. He received his BS degree from Hochiminh City University of Technology and Ph.D. degree from Kyung Hee University, in electrical and computer engineering, in 2005 and 2011, respectively. He was an Assistant Professor with Department of Computer Science and Engineering, Kyung Hee University, Korea from 2012 to 2017. His research interest is to applying analytic techniques of optimization, game theory, and stochastic modeling to cutting-edge applications such as cloud and mobile edge computing, data centers, heterogeneous wireless networks, and big data for networks. He received the best KHU thesis award in engineering in 2011 and best paper award at IEEE ICC 2016. He has been the Editor of IEEE Transactions on Green Communications and Networking since 2016, and served as the Editor of the 2017 Newsletter of Technical Committee on Cognitive Networks on Internet of Things.
\end{IEEEbiography}
\begin{IEEEbiography}[{\includegraphics[width=1in,height=1.25in,clip,keepaspectratio]{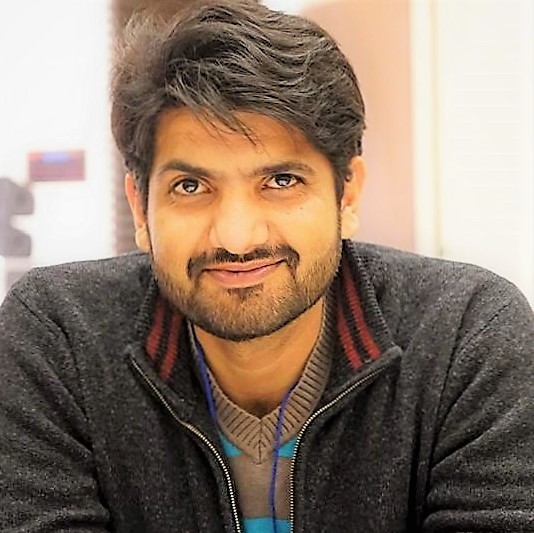}}]{S. M. Ahsan Kazmi} is with the Institute of Information Systems (IIS), Innopolis University, Innopolis, Tatarstan, Russia, where he is currently working as an Assistant Professor. He received his Master degree in Communication System Engineering from the National University of Sciences and Technology (NUST), Pakistan, in 2012, and the Ph.D. degree in Computer Science and Engineering from Kyung Hee University (KHU), South Korea. His research interests include applying analytical techniques of optimization and game theory to radio resource management for future cellular networks.
\end{IEEEbiography}
\begin{IEEEbiography}[{\includegraphics[width=1in,height=1.25in,clip,keepaspectratio]{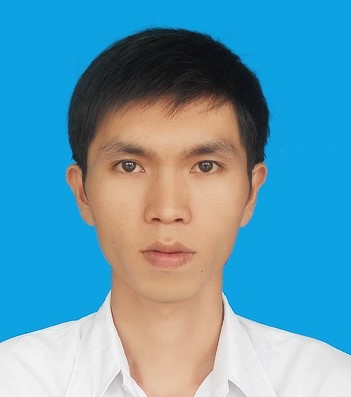}}]{Tri Nguyen Dang} is currently pursuing his Ph.D. in Computer Science and Engineering under fully funded scholarship at Kyung Hee University, South Korea.
He received his B.S. in Information Teaching from Hue University’s College of Education, Vietnam in 2014. His professional experience include Mobile Application Developer, middle-ware programming. He had won consolation prize of 
Olympiad in Informatics ACM ICPC Vietnam in 2013. His research interest includes Network Optimization, and Mobile Cloud computing, Mobile Edge Computing, Internet of Things.  
\end{IEEEbiography}
\begin{IEEEbiography}[{\includegraphics[width=1in,height=1.25in,clip,keepaspectratio]{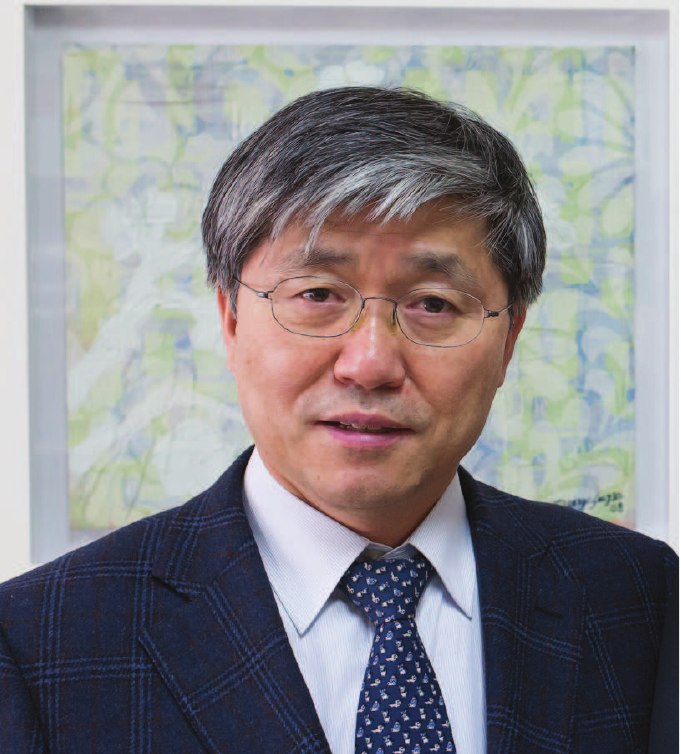}}]{Choong Seon Hong} (S’95-M’97-SM’11) received the B.S. and M.S. degrees in electronic engineering from Kyung Hee University, Seoul, South Korea, in 1983 and 1985, respectively, and the Ph.D. degree from Keio University, Minato, Japan, in 1997. In 1988, he joined Korea Telecom, where he worked on broadband networks as a Member of Technical Staff. In September 1993, he joined Keio University. He worked for the Telecommunications Network Laboratory, Korea Telecom, as a Senior Member of Technical Staff and the Director of the Networking Research Team until August 1999. Since September 1999, he has been a Professor with the Department of Computer Science and Engineering, Kyung Hee University. His research interests include future Internet, ad hoc networks, network management, and network security. He is a member of ACM, IEICE, IPSJ, KIISE, KICS, KIPS, and OSIA. He has served as the General Chair, a TPC Chair/Member, or an Organizing Committee Member for international conferences such as NOMS, IM, APNOMS, E2EMON, CCNC, ADSN, ICPP, DIM, WISA, BcN, TINA, SAINT, and ICOIN. In addition, he is currently an Associate Editor of the IEEE Transactions on Network and Service Management, International Journal of Network Management, and Journal of Communications and Networks and an Associate Technical Editor of the IEEE Communications Magazine. 
\end{IEEEbiography}
\balance
\end{document}